\documentclass[11pt,fleqn]{article}
\usepackage{amsmath,amssymb,graphicx,epsfig}

\def\bea{\begin{eqnarray}}
\def\eea{\end{eqnarray}}
\def\be{\begin{equation}}
\def\ee{\end{equation}}
\def\ba{\begin{array}}
\def\ea{\end{array}}
\def\nn{\nonumber}
\def\a{& \hspace{-11pt}}
\def\b{& \hspace{-7pt}}

\font\tenrsfs=rsfs10
\font\sevenrsfs=rsfs7
\font\fiversfs=rsfs5
\newfam\rsfsfam
\textfont\rsfsfam=\tenrsfs
\scriptfont\rsfsfam=\sevenrsfs
\scriptscriptfont\rsfsfam=\fiversfs
\def\mathscr#1{{\fam\rsfsfam\relax#1}}

\begin{document}

\thispagestyle{empty}

\begin{center}

$\;$

\vspace{1cm}

{\huge \bf Sequestering by global symmetries \\[2mm] in Calabi-Yau string models}

\vspace{1.5cm}

{\Large {\bf Christopher Andrey} and {\bf Claudio~A.~Scrucca}}\\[2mm] 

\vspace{0.6cm}

{\large \em Institut de Th\'eorie des Ph\'enom\`enes Physiques\\ 
Ecole Polytechnique F\'ed\'erale de Lausanne\\ 
CH-1015 Lausanne, Switzerland\\}

\vspace{0.2cm}

\end{center}

\vspace{1cm}

\centerline{\bf \large Abstract}
\begin{quote}

We study the possibility of realizing an effective sequestering between 
visible and hidden sectors in generic heterotic string models, generalizing 
previous work on orbifold constructions to smooth Calabi-Yau compactifications. 
In these theories, genuine sequestering is spoiled by interactions mixing chiral 
multiplets of the two sectors in the effective K\"ahler potential. These effective 
interactions however have a specific current-current-like structure and can 
be interpreted from an $M$-theory viewpoint as coming from the exchange 
of heavy vector multiplets. One may then 
attempt to inhibit the emergence of generic soft scalar masses in the visible 
sector by postulating a suitable global symmetry in the dynamics of the hidden 
sector. This mechanism is however not straightforward to implement, because 
the structure of the effective contact terms and the possible global symmetries 
is a priori model dependent. To assess whether there is any robust and generic 
option, we study the full dependence of the K\"ahler potential on the moduli and 
the matter fields. This is well known for orbifold models, where it always 
leads to a symmetric scalar manifold, but much less understood for Calabi-Yau 
models, where it generically leads to a non-symmetric scalar manifold. 
We then examine the possibility of an effective sequestering by global 
symmetries, and argue that whereas for orbifold models this can be put at 
work rather naturally, for Calabi-Yau models it can only be implemented 
in rather peculiar circumstances. 

\vspace{5pt}
\end{quote}

\renewcommand{\theequation}{\thesection.\arabic{equation}}

\newpage

\setcounter{page}{1}

\section{Introduction}
\setcounter{equation}{0}

In supergravity models, it is natural to imagine that supersymmetry breaking 
occurs at an intermediate scale in a hidden sector and is dominantly mediated 
to the visible sector by gravitational interactions, with the net effect of inducing soft 
breaking terms of a size close to the electroweak scale. These soft terms are 
however induced through higher-dimensional operators mixing visible and hidden 
sector fields in the effective theory, with a structure that depends on the details of 
the underlying microscopic  theory and is therefore a priori generic. In particular, 
one naturally expects soft scalar masses with a generic flavor structure, while the 
non-observation of certain flavor changing processes instead requires these to be 
approximately universal. This leads to the so-called supersymmetric flavor problem, 
which consists in finding a natural and robust explanation for the approximate flavor 
universality that soft scalar masses need to enjoy. 

One of the most interesting proposals for solving this problem is the idea of 
sequestering the visible and the hidden sectors by localizing them on two 
distinct branes at different positions along an extra dimension \cite{seq}.
In the basic situation where these two sectors interact only through minimal 
gravity in the bulk, which corresponds to the so-called no-scale models 
\cite{noscale}, local contact terms between the two brane sectors are 
guaranteed to be absent. Moreover, contact terms between each brane 
sector and the additional radion chiral multiplet arising in the bulk, which 
can also participate to supersymmetry breaking, turn out to be absent too.
As a consequence, scalar masses vanish at the classical level and are 
induced only by non-local loop effects of various kinds, like for instance 
anomaly mediation \cite{seq,anomed}, radion-mediation \cite{GR} or 
brane-to-brane mediation \cite{RSS,BGGLLNP}, which have the crucial 
common characteristic of being approximately flavor-universal. Thanks 
to this property, this minimal setup allows to construct phenomenologically 
acceptable and satisfactory effective models based on 5D supergravity 
theories with one compact dimension.

In string models, which are supposed to be the microscopic theories underlying
supergravity models, the framework that is needed to implement sequestering
arises very naturally, since the emergence of extra dimensions and localized 
matter sectors is almost unavoidable. It has however been emphasized in 
\cite{noseq} that there is an endemic difficulty in realizing the minimal setup 
needed for sequestering. As a matter of fact, in most of the string models where 
the 4D low-energy effective theory has been worked out, there appear non-trivial 
contact terms between matter sectors in the effective K\"ahler potential, even when 
these are sequestered at distinct points in the internal compact space, as well as
couplings between each matter sector and the non-minimal moduli sector. As a result, 
non-vanishing and non-universal soft scalar masses generically arise at the classical 
level. From the perspective of the 5D intermediate effective theory obtained by retaining 
only the compact dimension separating the visible and the hidden sectors, these 
effects were interpreted in \cite{noseq} as being induced by additional vector multiplets 
propagating in the bulk and coupling non-minimally to the localized brane sectors. 
Since these vector multiplets appear very generically, one is then forced to conclude 
that sequestering is rather unnatural to realize in string models.

For heterotic models based on a compact manifold $X$ with a vector bundle $V$ over it
\cite{CYhet,Torhet}, the above phenomenon can be visualized very clearly. Indeed, these 
models have a simple interpretation within $M$-theory, where the additional extra dimension 
is a segment connecting two branes supporting charged sectors \cite{HW}. These two brane 
sectors are then naturally identified with visible and hidden sectors. In the weak-coupling 
regime, which corresponds to a small size for the extra segment, the heterotic and $M$-theory 
pictures becomes equivalent, the former being obtained by integrating out the heavy KK modes 
in the latter. After compactifying on $X$, this implies a similar relation between the 4D effective
theory and a 5D theory compactified on a segment connecting two 4D branes. In the bulk of this 
theory one obtains one vector multiplet for each non-trivial K\"ahler structure deformation 
of $X$, with couplings to the brane sectors that are determined by the choice of $V$ 
through a shift in its Bianchi identity \cite{CCDF,AFT,LOSW}. From the 4D point of view, 
each of these multiplets contains one chiral multiplet zero mode describing a non-universal 
modulus of $X$ in the low-energy effective theory and one tower of vector multiplet KK 
modes inducing non-trivial effective interactions when integrated out. The non-trivial contact 
terms of the 4D effective K\"ahler potential are then in one-to-one correspondence with 
the presence of non-minimal K\"ahler moduli for $X$, besides the one controlling its 
overall volume, and have a structure that depends on the choice of $V$.

For orientifold models based on a compact manifold $X$ with $D$-branes wrapped on it 
(see for example \cite{CYor1,CYor2} for recent reviews), the situation is similar. Visible and hidden 
sectors may naturally arise from $D$-branes wrapping on two non-intersecting cycles of $X$. 
It is however less straightforward to relate the 4D effective theory to a higher-dimensional 
theory and reinterpret the contact terms as being induced by the exchange of heavy fields. 
Nonetheless, it turns out that in all the cases where it has been worked out, the 4D effective 
theory displays a structure that is very similar to the one arising in heterotic models. In particular, 
the contact terms arising in the 4D effective K\"ahler potential again seem to be in one-to-one 
correspondence with non-minimal K\"ahler moduli of $X$, suggesting that in this case too one 
should be able to interpret these as due to the exchange of corresponding heavy vector multiplets. 
A precise argumentation justifying this conclusion was presented in \cite{noseq} for the special 
case of toroidal orientifolds, where one can make use of $T$-duality to reach a situation where 
the two sectors are again separated by a single extra dimension, and it is plausible that it indeed 
holds more in general.

In summary, we see that in string models one may naturally achieve the situation 
where the visible and hidden sectors are split along an extra dimension, but this is 
not enough to really achieve sequestering. Nevertheless, the situation is still better 
that in a generic supergravity model, because the non-vanishing contact terms that 
arise in the K\"ahler potential have a very specific form, as a consequence of the fact 
that they are induced by the exchange of heavy vector multiplets. More precisely, 
these contact terms consist of products of two or more of the current superfields 
$J^a_{\rm v}$ and $J^a_{\rm h}$ that act as sources for the heavy vector superfields. 
One may then hope to be able to exploit the structure of these classical 
contact terms to devise situations where they actually give a satisfactory contribution 
to soft masses. In playing this game, one may take the point of view of \cite{KL,BIM} 
that the effective K\"ahler potential, which controls through the contact interactions mixing 
visible and hidden sectors the general structure of soft scalar masses, is known and 
therefore fixed, whereas the superpotential, which controls the size of the supersymmetry 
breaking auxiliary fields of the hidden sector fields, is not known and a priori generic. 
For generality, one should moreover consider the situation where both the moduli fields 
and the hidden brane fields participate to supersymmetry breaking. Finally one may also 
take into account the fact that there are constraints from the condition that the supersymmetry 
breaking sector should admit a metastable de Sitter vacuum with sufficiently small energy 
and sufficiently long life time. For a given K\"ahler potential, this constrains the 
acceptable directions for the Goldstino vector of auxiliary fields and therefore the acceptable 
superpotentials \cite{GRS}.

A first appealing possibility is to assume that the moduli fields dominate supersymmetry 
breaking and that for some reason the contact terms between these fields and the visible 
sector fields are flavor universal \cite{KL,BIM}. In that case one would get a non-vanishing 
but flavor-universal classical contribution to soft scalar masses, and loop contributions would 
only represent a small correction. This scenario would for instance naturally occur if the 
dilaton could dominate supersymmetry breaking on its own, since its couplings are 
automatically universal at the classical level \cite{KL}. But unfortunately, it turns out that 
due to the leading order form of the K\"ahler potential for the dilaton, the assumption that it 
dominates supersymmetry breaking is actually incompatible with the existence of a metastable 
de Sitter vacuum, at least under the assumption that the string coupling is weak \cite{C,BdA,GRS}. 

Another appealing possibility is to imagine that the hidden brane fields dominate supersymmetry 
breaking and that their dynamics enjoys a set of global symmetries ensuring the conservation of 
the hidden-sector current superfields $J^a_{\rm h}$, which appear together with visible-sector 
current superfields $J^a_{\rm v}$ in the contact terms \cite{KMS}. In such a situation, the classical 
contribution to the soft scalar masses would cancel out, at least at leading order in the hidden scalar 
expectation values, and flavor-universal loop corrections would represent the dominant effect.
The basic point behind this idea was already explained in \cite{SS}, although in a different
context and in the approximation of rigid supersymmetry, and rests on the fact that the conservation 
of the superfields $J^a_{\rm h}$ implies that both their $F$ and $D$ components vanish. The consequent
vanishing of classical soft scalar masses can then also be viewed as a cancellation between 
the various contributions coming from the hidden sector fields, which is determined by the 
constraints put on the ratios of their auxiliary fields by the invariance of the superpotential 
under the global symmetries. In our previous paper \cite{ASmild}, we studied how this nice 
framework may be implemented in supergravity models. We showed that the cancellation 
mechanism is generically spoiled by non-linear effects coming from terms with more than 
two currents in the contact interactions, as well as by gravitational effects in the Ward identity 
of the global symmetries. We however also argued that both of these effects become small 
in the limit of small expectation values for the hidden-sector matter scalar fields, and can in practice 
be safely neglected. In this situation, one would thus recover a milder form of the sequestering 
mechanism, working thanks to global symmetries. 

The aim of this work is to understand whether it is be possible to implement the above mechanism 
of sequestering by global symmetries within generic string models and with both the moduli and the 
matter fields participating to supersymmetry breaking. More specifically, we want to 
clarify the circumstances under which it is possible to find suitable global symmetries ensuring the 
conservation of the currents building up the contact terms. In fact, it is a priori not automatic that such 
symmetries exist, because the couplings of the heavy vector multiplets to the brane fields are not 
minimal gauge couplings, but rather dictated by modified Bianchi identities, and one may then 
wonder how natural it is that they arise. For concreteness and simplicity, we shall focus on the case 
of heterotic models, but we expect that it should be possible to perform a similar study for orientifold 
models. In \cite{ASmild} we examined the special subclass of models based on orbifolds, and found 
that in that case the needed global symmetries naturally arise. Our goal here is to study what 
happens in the more general case of models based on smooth Calabi-Yau manifolds, and in 
particular whether the needed global symmetries still arise in a natural way. One major difficulty in 
this generalization concerns the knowledge of the effective K\"ahler potential. For orbifold models, 
the exact dependence on both the moduli fields and the brane fields is known \cite{Korb1,Korb2},
and the structure of contact terms is therefore well under control.
For generic Calabi-Yau manifolds, on the other hand, only the dependence on the moduli fields is 
known exactly \cite{Strominger,CFG,CdlO}, whereas the knowledge of the dependence on the brane 
fields is mostly limited to the leading quadratic order \cite{DKL}. An interesting claim on the structure 
of the exact dependence on the matter fields has however recently appeared in the literature, based 
on the higher-dimensional $M$-theory interpretation of these models \cite{PS}. This generalizes the result 
of \cite{W} applying to the special case of Calabi-Yau manifolds possessing only a minimal volume 
K\"ahler modulus. It moreover has a structure that is qualitatively similar to the one derived in 
\cite{Korient1,Korient2,Korient3} for orientifold models. One of our main tasks will then be to 
assess this result from the standard heterotic string point of view and to study the resulting structure 
of contact terms.

The rest of the paper is organized as follows. In section 2 we consider the heterotic string compactified 
on a smooth Calabi-Yau manifold and study the general structure of the effective K\"ahler potential. 
In section 3 we consider similarly the heterotic string compactified on a toroidal orbifold and show 
how the effective K\"ahler potential for the untwisted sector can be understood in similar terms.
In section 4 we comment on the $M$-theory interpretation of these models and the way in which 
the contact terms arising in the effective K\"ahler potential can be understood as emerging from 
the exchange of heavy vector multiplets. In section 5 we study the scalar manifolds emerging 
in these models and discuss a canonical parametrization that is particularly convenient to describe 
the neighborhood of the reference point where only the universal volume modulus has a scalar 
expectation value. In section 6 we study the structure of soft scalar masses at this reference point
and examine under which circumstances they may be made to vanish by imposing some global 
symmetry. In section 7 we present our conclusions. Finally, in appendix A we summarize some 
basic facts about Calabi-Yau manifolds and vector bundles over them, and in appendix B we 
record some useful facts about the symmetric spaces emerging in orbifold models.

\section{The heterotic string on a Calabi-Yau manifold}
\setcounter{equation}{0}

Let us consider the heterotic string compactified on a generic Calabi-Yau manifold $X$ with a 
generic stable holomorphic vector bundle $V$ over it \cite{CYhet}. The 4D low-energy effective 
supergravity theory describing such a model can be obtained by starting from the 10D supergravity 
effective theory and working out its reduction on $X$. We shall start by reviewing the general 
structure of these models. We shall next describe how the effective K\"ahler potential can be 
derived by computing the form of the bosonic kinetic terms. We shall focus on the K\"ahler 
moduli and the matter fields, and study the full dependence of the K\"ahler potential on these 
fields, generalizing the known results for the exact dependence on the moduli and the leading 
dependence on the matter fields.

\subsection{General structure}

In the original 10D effective supergravity theory, the bosonic fields are the metric $G_{MN}$, the 
antisymmetric tensor $B_{MN}$, the dilaton $\Phi$ and the $E_8 \times E_8$ gauge fields $A_M^X$. 
It is convenient to describe $B_{MN}$ in terms of a $2$-form $B$ and $A_M^X$ in terms of 
a Lie-algebra-valued $1$-form $A$. At the two-derivative order, the effective action for these 
fields reads:
\bea
S_{10} \b=\b \frac 1{\kappa_{10}^2}\int \! d^{10}x \sqrt{-G} e^{- 2 \Phi} \bigg[\!-\! \frac 12 R 
+ 2\, \partial_M \Phi \, \partial^M \Phi - \frac 1{4} |H|^2 \bigg] \nn \\
\b\;\b + \frac 1{g_{10}^2}\int \! d^{10}x \sqrt{-G} e^{- 2 \Phi} \bigg[\!-\! \frac 12 {\rm tr} |F|^2\bigg] \,.
\label{10Dkin}
\eea
The 10D gravitational and gauge couplings $\kappa_{10}^2$ and $g_{10}^2$ are related to the 
string slope $\alpha'$ through the formula $\kappa_{10}^2/g_{10}^2 = \alpha'/4$. The $2$-form 
$F$ denotes the usual field-strength of the non-Abelian gauge field $A$ and the $3$-form 
$\Gamma$ the Chern-Simons form associated to it, whereas the $3$-form $H$ is a modified 
field-strength for the Abelian antisymmetric field $B$:
\bea
\a\a F = d A + A \wedge A \,,\;\; \Gamma = {\rm tr} \Big(A \wedge d A + \frac 23 A \wedge A \wedge A \Big) \,, \\
\a\a H = d B - \frac {\kappa_{10}^2}{g_{10}^2} \Gamma \,.
\eea
At higher order in the derivative expansion, there appear other terms involving the curvature $2$-form 
$R = d\omega + \omega \wedge \omega$ related to the spin connection $1$-form $\omega$, as well as the 
Chern-Simons $3$-form $\Xi = {\rm tr}(\omega \wedge d \omega + 2/3 \, \omega \wedge \omega \wedge \omega)$ 
associated to it. In particular, at the four-derivative level one gets extra terms that essentially correspond to substituting
${\rm tr} |F|^2$ with ${\rm tr} |F|^2 - {\rm tr} |R|^2$ and $\Gamma$ with $\Gamma - \Xi$. These two kinds 
of new terms are related by supersymmetry, and turn out to be relevant for the consistency of the microscopic theory. 
Most importantly, the Bianchi identity for the $3$-form $H$ becomes
\be
d H =  \frac {\kappa_{10}^2}{g_{10}^2} \Big({\rm tr} (R \wedge R) - {\rm tr} (F \wedge F)\Big) \,.
\label{bianchi}
\ee
Consistent supersymmetric backgrounds must not only lead to vanishing supersymmetry transformations of 
the fermions, but also solve the above Bianchi identity. In particular, the right-hand side of (\ref{bianchi}) must 
vanish in cohomology. This represents a topological relation between the tangent bundle $TX$ of the 
compactification manifold $X$ and the vector bundle $V$ over it, which restricts the possible choices of 
$V$ for a given $X$. One simple and universal possibility, called standard embedding, is to take $V$ to 
be isomorphic to $TX$. This means that $V$ has structure group $SU(3)$ and that the background 
values of the gauge connection $A$ and the spin connections $\omega$ are identified. In such a case the 
right-hand side of (\ref{bianchi}) vanishes identically and the background is a Calabi-Yau geometry for $G_{MN}$. 
A more general possibility, called non-standard embedding, is to require that $V$ should have the 
same second Chern character as $TX$. This allows $V$ to have more general structure groups and the 
background values of $A$ and $\omega$ to differ \cite{Nonstand1,Nonstand2}. In this more general case, 
however, the right-hand side of (\ref{bianchi}) does not vanish identically but only modulo an exact form. 
As a result, the background is no-longer a simple Calabi-Yau geometry for $G_{MN}$ and also involves a 
non-trivial profile for $B_{MN}$ and $\Phi$. However, it has been argued in \cite{Nonstand1} that such a 
background exists and that it can be understood as a deformation of the standard case in a large volume or 
small $\alpha'$ expansion. Some of the leading corrections have been worked out in \cite{Corr1,Corr2,Corr3}. 

To characterize the models resulting from this construction, one can start by classifying the relevant modes in 
terms of representations under the holonomy group $SU(3)$ of $X$ and the structure group $S$ of $V$. 
The 10D Lorentz group $SO(1,9)$ is broken to \mbox{$SO(1,3) \times U(1) \times SU(3)$}, where the $SO(1,3)$ factor 
survives as 4D Lorentz symmetry. The fundamental representation splits as ${\bf 10} \to {\bf 4} \oplus {\bf 3} \oplus {\bf \bar 3}$. 
We correspondingly split the 10D Lorentz indices $M$ into 4D Lorentz indices $\mu$ and internal $SU(3)$ indices 
$i,\bar \imath$. The 10D gauge group $E_8 \times E_8$ is broken to $G \times S$, where $G$ is the commutant 
of $S$ and survives as 4D gauge group. One actually gets $S = S_{\rm v} \times S_{\rm h}$ and 
$G = G_{\rm v} \times G_{\rm h}$, where $G_{\rm v}$ and $G_{\rm h}$ are the commutants of $S_{\rm v}$ 
and $S_{\rm h}$ within the two $E_8$ factors, but for the moment we shall treat the two sectors together. 
The adjoint representation splits pretty generically as ${\bf 496} \to ({\bf Adj}, {\bf 1}) \oplus ({\bf 1}, {\bf Adj}) 
\oplus ({\bf R}, {\bf r}) \oplus ({\bf \bar R}, {\bf \bar r})$, where ${\bf R}$ and ${\bf r}$ are complex and generically 
reducible representations of $G$ and $S$ (except for a few special cases that we shall disregard for notational 
simplicity). We correspondingly split the 10D gauge indices $X$ of the adjoint representation of $E_8 \times E_8$ into 
4D gauge indices $x$ in the adjoint representation of $G$, indices $\rho$ in the adjoint representation of $S$ 
and indices $\alpha \epsilon$ and $\bar \alpha \bar \epsilon$ in the representations that are left over. Using the 
above decompositions, one may now classify the bosonic fields in terms of representations of $SU(3) \times S$. 
In the neutral sector, the fields transform non-trivially only under $SU(3)$ but are all singlets under $S$.
$G_{\mu \nu}$ gives a symmetric tensor in the ${\bf 1}$, $G_{\mu i}$, $G_{\mu \bar \imath}$ give
vectors in the ${\bf 3} \oplus {\bf \bar 3}$, and $G_{i \bar \jmath}$, $G_{ij}$, $G_{\bar \imath \bar \jmath}$ 
give scalars in the ${\bf 1} \oplus {\bf 8} \oplus {\bf 6} \oplus {\bf \bar 6}$. Similarly $B_{\mu \nu}$ 
gives an antisymmetric tensor dual to a scalar in the ${\bf 1}$, $B_{\mu i}$, $B_{\mu \bar \imath}$ give 
vectors in the ${\bf 3} \oplus {\bf \bar 3}$, and $B_{i \bar \jmath}$, $B_{ij}$, $B_{\bar \imath \bar \jmath}$ 
give scalars in the ${\bf 1} \oplus {\bf 8} \oplus {\bf 3} \oplus {\bf \bar 3}$. Finally $\Phi$ gives just a
scalar in the ${\bf 1}$. In the charged sector, on the other hand, the fields transform non-trivially under 
$SU(3) \times S$.  $A_\mu^x$ gives vectors in the $({\bf 1},{\bf 1})$, $A_\mu^{\rho}$ gives
vectors in the $({\bf 1},{\bf Adj})$, $A_\mu^{\alpha\epsilon}$ and $A_\mu^{\bar \alpha \bar \epsilon}$ 
give vectors in the $({\bf 1},{\bf r}) \oplus ({\bf 1},{\bf \bar r})$, $A_i^x$, 
$A_{\bar \imath}^x$ give scalars in the $({\bf 3},{\bf 1}) \oplus ({\bf \bar 3},{\bf 1})$, 
$A_i^{\rho}$, $A_{\bar \imath}^{\rho}$ give scalars in the $({\bf 3},{\bf Adj}) \oplus ({\bf \bar 3},{\bf Adj})$, 
and finally $A_i^{\alpha \epsilon}$, $A_{\bar \imath}^{\alpha \epsilon}$,
$A_i^{\bar \alpha \bar \epsilon}$, $A_{\bar \imath}^{\bar \alpha \bar \epsilon}$
give scalars in the $({\bf 3},{\bf r}) \oplus ({\bf \bar 3},{\bf r}) \oplus ({\bf 3},{\bf \bar r}) \oplus ({\bf \bar 3},{\bf \bar r})$.

The spectrum of light fields entering the 4D low-energy effective theory is determined by figuring out all the 
zero-modes admitted by the above 10D bosonic fields. This is done by associating these modes to differential 
forms on $X$ taking values in appropriate vector bundles constructed out of $TX$ or $V$, and looking for all the 
possible independent harmonic components of these forms. The linear space of such harmonic forms is known 
to be in one-to-one correspondence with non-trivial equivalence classes of the Dolbeault cohomology groups,
and one may then use bases of such spaces to parametrize the various light fields. For neutral fields, what matters 
are the tangent and cotangent bundles $TX$ and $T^*X$ with structure group $SU(3)$, and the components 
of the relevant harmonic forms fill representations of $SU(3)$. There is $1$ harmonic $(3,0)$ form $\Omega$ 
and its conjugate filling the ${\bf 1} \oplus {\bf 1}$, 
$h^{1,2} = {\rm dim} (H^1(X,TX))$ harmonic $(1,2)$ forms $\sigma_Z$ filling the 
${\bf 6} \oplus {\bf \bar 6}$, and finally $h^{1,1} = {\rm dim} (H^1(X,T^*\!X))$ harmonic $(1,1)$ forms 
$\omega_A$ filling the ${\bf 1} \oplus {\bf 8}$.
For charged fields, what matter are the bundles $V_{\rm Adj}$, $V_{\rm r}$ and $V_{\rm \bar r}$ obtained 
by lifting $V$ to the representations ${\bf Adj}$, ${\bf r}$ and ${\bf \bar r}$ of its structure group $S$, and 
the components of the relevant harmonic forms fill representations of $SU(3) \times S$. There are 
$n_{1} = {\rm} {\rm dim} (H^{1}(X,V_{\rm Adj}))$ harmonic $(1,0)$ forms $\sigma_\Theta$ and their conjugate 
filling the $({\bf 3},{\bf Adj}) \oplus ({\bf \bar 3},{\bf Adj})$, 
$n_{R} = {\rm} {\rm dim} (H^{1}(X,V_{\rm r}))$ harmonic $(1,0)$ forms $u_P$ and their conjugate filling the 
$({\bf 3},{\bf r}) \oplus ({\bf \bar 3},{\bf \bar r})$, and $n_{\bar R} = {\rm} {\rm dim} (H^{1}(X,V_{\rm \bar r}))$ 
harmonic $(1,0)$ forms $v_K$ and their conjugate filling the $({\bf 3},{\bf \bar r}) \oplus ({\bf \bar 3},{\bf r})$. 
Using the above set of harmonic forms, one finally finds the following spectrum of light 4D bosonic fields.
In the neutral sector, there is $1$ symmetric tensor coming from $G_{\mu \nu}$ and belonging to the gravitational 
multiplet $G$, $1$ universal complex scalar coming from $\Phi$ and $B_{\mu\nu}$ and belonging to the dilaton 
chiral multiplet $S$, $h^{1,1}$ complex scalars coming from the decomposition of the forms associated to 
$G_{i \bar \jmath}$ and $B_{i \bar \jmath}$ onto the basis $\omega_A$ and belonging to K\"ahler moduli chiral 
multiplets $T^A$, and finally $h^{1,2}$ complex scalars coming from the decomposition of the forms associated 
to $G_{ij}$ and $G_{\bar \imath \bar \jmath}$ onto the basis $\sigma_Z$ and belonging to complex structure moduli 
chiral multiplets $U^Z$. In the charged sector, there is $1$ set of vectors coming from $A_\mu^x$ and belonging 
to vector multiplets $V^x$ in the ${\bf Adj}$ of $G$, $n_{1}$ complex scalars coming from $A_i^{\rho}$ and $A_{\bar \imath}^{\rho}$ 
and belonging to vector bundle moduli chiral multiplets $E^\Theta$ in the ${\bf 1}$ of $G$, $n_{R}$ sets of complex scalars 
coming from $A{}_i{\hspace{-3pt}\raisebox{8pt}{$$}}^{\alpha \xi}$ and $A{}_{\bar \imath}{\hspace{-3pt}\raisebox{8pt}{$$}}^{\alpha\xi}$ 
and belonging to matter chiral multiplets $\Phi^{P \alpha}$ in the ${\bf R}$ of $G$, and finally $n_{\bar R}$ sets of 
complex scalars coming from $A{}_i{\hspace{-3pt}\raisebox{8pt}{$$}}^{\bar \alpha \bar \xi}$ and
$A{}_{\bar \imath}{\hspace{-3pt}\raisebox{8pt}{$$}}^{\bar \alpha \bar \xi}$ and belonging to matter 
chiral multiplets $\Psi^{K \bar \alpha}$ in the ${\bf \bar R}$ of $G$.

The models with the simplest gauge quantum numbers are obtained by choosing bundles whose structure 
group involves factors that are either trivial or equal to $SU(3)$ in each of the two sectors. 
In the first case one has $E_8 \to E_8$ with ${\bf 248} \to {\bf 248}$, and the gauge group in unbroken.
In the second case one has $E_8 \to E_6 \times SU(3)$ with ${\bf 248} \to ({\bf 78},{\bf 1}) \oplus ({\bf 1},{\bf 8}) 
\oplus ({\bf 27},{\bf 3}) \oplus ({\bf \overline{27}},{\bf \overline 3})$, but nothing from the $SU(3)$ factor survives 
and the gauge group is thus broken to $E_6$.
A first type of model can be constructed by making the asymmetric choice $S_{\rm v} = SU(3)$, $S_{\rm h} = \text{trivial}$.
One then finds $G_{\rm v} = E_6$ and $n_1^{\rm v} = {\rm dim} (H^1(X, V_{\rm v} \otimes V^*_{\rm v}))$, 
$n_{27}^{\rm v} = {\rm dim} (H^1(X,V_{\rm v}))$, $n_{\overline{27}}^{\rm v} = {\rm dim} (H^1(X,V^*_{\rm v}))$ in 
the visible sector, and just $G_{\rm h} = E_8$ in the hidden sector. The standard embedding where $V$ is isomorphic
to $TX$ is a particular case of this class of models where the Bianchi identity is automatically 
satisfied, with the special property that $n_{27} = h^{1,2}$ and $n_{\overline{27}} = h^{1,1}$. A second type of model 
can be constructed by making the symmetric choice $S_{\rm v} = SU(3)$, $S_{\rm h} = \text{SU(3)}$.
One then finds $G_{\rm v} = E_6$ and $n_1^{\rm v} = {\rm dim} (H^1(X, V_{\rm v} \otimes V^*_{\rm v}))$,
$n_{27}^{\rm v} = {\rm dim} (H^1(X,V_{\rm v}))$, $n_{\overline{27}}^{\rm v} = {\rm dim} (H^1(X,V^*_{\rm v}))$
in the visible sector and similarly $G_{\rm h} = E_6$ and $n_1^{\rm h} = {\rm dim} (H^1(X, V_{\rm h} \otimes V^*_{\rm h}))$, 
$n_{27}^{\rm h} = {\rm dim} (H^1(X,V_{\rm h}))$, $n_{\overline{27}}^{\rm h} = {\rm dim} (H^1(X,V^*_{\rm h}))$ in 
the hidden sector. Notice that in this case $V_{\rm v}$ and $V_{\rm h}$ are not allowed to be isomorphic to 
$TX$, because this would violate the Bianchi identity.

\subsection{Effective K\"ahler potential}

The 4D effective K\"ahler potential can be determined by performing the reduction of the 10D effective 
kinetic terms for the bosonic fields by integrating over the compact Calabi-Yau $X$ and comparing the result 
with the standard general form of the Lagrangian of 4D supergravity theories. To perform this computation, 
we will make two approximations which are commonly done and which crucially simplify the task. The first 
approximation is that we will ignore the higher-derivative corrections to the 10D effective action and the 
deformations of the background, and therefore simply consider the reduction of the action (\ref{10Dkin}) 
onto a generic Calabi-Yau $X$ with a generic stable holomorphic vector bundle $V$ over it. This implies 
that the result will only be accurate for terms involving arbitrary powers of the moduli fields and arbitrary 
powers of the combination of $\alpha'$ times two matter fields, and will miss corrections involving powers 
of $\alpha'$ that are not accompanied by two matter fields, but this is not a big limitation for our purposes. 
The second approximation is that we will ignore the effect of properly integrating out massive Kaluza-Klein 
modes and restrict to the truncation of the action to the 4D massless zero-modes. This would generically 
imply that the result is accurate only for terms involving an arbitrary number of moduli but at most two matter 
fields, since terms with four and more matter fields can receive corrections induced by the exchange of 
heavy neutral modes, and this would represent a dramatic limitation for our purposes. We will therefore 
imagine to restrict to those models for which these effects happen to be absent, at least for the term 
involving four matter fields in which we are primarily interested. This is guaranteed to happen if there
is no cubic coupling between two light matter modes and one heavy moduli mode. 
Finally, we shall for simplicity restrict our attention to the $h^{1,1}$ K\"ahler moduli and the $n_{R}$ 
families of charged matter fields in the representation ${\bf R}$, and instead completely discard the 
dilaton, the $h^{1,2}$ complex structure moduli, the $n_{1}$ vector bundle moduli and the $n_{\bar R}$ 
families of matter fields in the representation ${\bf \bar R}$.

To compute the 4D effective kinetic terms, we now proceed as follows. We start from eq.~(\ref{10Dkin})
restricted to the modes associated to $G_{i \bar \jmath}$, $B_{i \bar \jmath}$ and $A_i$ and integrate 
over the internal manifold $X$. We then express the result in terms of the 4D gravitational and gauge 
couplings. These are defined as $\kappa_4^2 = \kappa_{10}^2/V$ and $g_4^2 = g_{10}^2/V$, where 
$V$ denotes the background value of the volume of the manifold $X$, and are again related as 
$\kappa_4^2/g_4^2 = \alpha'/4$. In the following, we shall set $\kappa_4 = 1$ by a choice of units. 
Moreover we shall effectively set $g_4 = 1$ in the scalar sector of the Lagrangian by suitably 
rescaling the charged matter fields. This corresponds to setting $\alpha' = 4$. In this way,
one finds the following result:
\bea
{\cal L}_{4} \b=\b \frac 1V \! \int \!\! d^6y \sqrt{G} \bigg[ 
\!-\! \frac 14 G^{i \bar \jmath} G^{p \bar q} \, \partial_\mu G_{i \bar q} \partial^\mu G_{p \bar \jmath} \nn \\
\b\;\b \hspace{61pt} +\, \frac 14 G^{i \bar \jmath} G^{p \bar q} \Big(\partial_\mu B_{i \bar q} + {\rm tr} 
(A_i \raisebox{8pt}{\footnotesize $\leftrightarrow$}\hspace{-9pt} \partial_\mu \bar A_{\bar q})\Big)
\Big(\partial^\mu B_{p \bar \jmath} + {\rm tr} 
(A_p \raisebox{8pt}{\footnotesize $\leftrightarrow$}\hspace{-9pt} \partial^\mu \! \bar A_{\bar \jmath})\Big) \nn \\[1mm]
\b\;\b \hspace{61pt} -\, G^{i \bar \jmath} \, {\rm tr} (\partial_\mu A_i \partial^\mu \! \bar A_{\bar \jmath}) 
\bigg] \,.
\label{kinterms}
\eea
To proceed, we associate the 10D fields $G_{i \bar \jmath}$, $B_{i \bar \jmath}$ and $A_i$ to differential 
forms $J$, $B$ and $A$, which are defined as follows in local complex coordinates $z^i$:
\bea
\a\a J = i G_{i \bar \jmath}\, dz^i \wedge d\bar z^{\bar \jmath} \,, \\
\a\a B = B_{i \bar \jmath} \, d z^i \wedge d \bar z^{\bar \jmath} \,, \\
\a\a A = A_i dz^i \,.
\eea
We then decompose these forms onto suitable bases of harmonic forms, with coefficients identified with the 
4D light fields. Some basic notation and results concerning harmonic forms on compact Calabi-Yau manifolds 
$X$ and stable holomorphic bundles $V$ over them are recorded for convenience in appendix A. 
To define the moduli fields, we shall need to introduce a basis of harmonic $(1,1)$ forms 
$\omega_A = \omega_{A i \bar \jmath} \, dz^i \wedge d \bar z^{\bar \jmath}$ on $X$,
which can also be viewed as $1$ forms with values in $T^*X$ over $X$:
\be
\omega_A \,,\;\; A = 0,\cdots,h^{1,1}-1 
\,:\;\;\text{basis of}\;\; H^{1,1}(X) \simeq H^1(X,T^*X) \,.
\ee
To define the matter fields, we shall also need a basis of Lie-algebra-valued harmonic $1$-forms 
$u_P = u_{P i}\, d z^i$ on $V_r$ over $X$:
\be
u_P \,,\;\; P=1,\cdots,n_R \,:\;\; \text{basis of}\;\; H^1(X,V_r) \,.
\ee
We observe now that the forms constructed by taking the product of one $u_P$ and one conjugate $\bar u_Q$ 
and tracing over the representation ${\bf r}$ yield $(1,1)$ forms on $X$. These $(1,1)$ forms are related to the 
description of the gauge invariant composite field that can be formed out of two charged matter fields. Since 
they play an important role, we shall define a dedicated symbol for them:
\be
c_{PQ} = i\, {\rm tr} (u_P \wedge \bar u_Q) \,:\;\; \text{generic $(1,1)$ forms on $X$} \,.
\label{deflambda}
\ee
A crucial observation is that these $(1,1)$ forms are however generically not harmonic. As a result, their 
scalar product with the non-harmonic $(1,1)$ forms describing massive neutral modes is in general non-vanishing.

It turns out that the low-energy effective K\"ahler potential always depends on the volume $V$ of $X$, 
which is given by the following expression in terms of the K\"ahler form $J$:
\be
V = \frac 16 \int_X \! J \wedge J \wedge J \,.
\ee
More explicitly, when rewritten in terms of the 4D fields describing the moduli and matter fields, this 
will depend on two quantities characterizing $X$ and $V$. The first is given by the integral of three harmonic 
$(1,1)$ forms $\omega_A$, $\omega_B$ and $\omega_C$, which defines the intersection numbers of $X$:
\be
d_{ABC} = \int_X \! \omega_A \wedge \omega_B \wedge \omega_C \,.
\label{dABC}
\ee
The second is given by the integral of the $(1,1)$ forms $c_{PQ}$ and a dual harmonic $(2,2)$
form $\omega^A$, which defines the component of the harmonic part of $c_{PQ}$ along $\omega_A$
and encodes therefore the overlap between the traced product of the $1$-forms $u_P$ and $\bar u_Q$ with 
the $(1,1)$ forms $\omega_A$:
\be
c^A_{PQ} = \int_X \! \omega^A \wedge c_{PQ} \,.
\label{lambdaAST}
\ee
It should be emphasized that (\ref{dABC}) is a topological invariant, as a result of the fact that the forms $\omega_A$ 
are harmonic, whereas (\ref{lambdaAST}) is a priori not, because the forms $c_{ST}$ are in general not harmonic.

In the following, we shall restrict to the special case where the forms $c_{PQ}$ are harmonic and 
$c^A_{PQ}$ is a constant topological invariant, and derive the low-energy effective K\"ahler potential 
under these assumptions. We believe that this is a priori necessary to guarantee that the result obtained 
by truncating to the massless modes, without properly integrating out the massive modes, is reliable.
But as matter of fact, we will also crucially exploit these assumptions to be able to obtain a simple result.
We shall discuss in subsection 2.3 what may happen in the more general case where $c_{PQ}$ is not 
harmonic and $c^A_{PQ}$ is not a topological invariant. For notational simplicity, we shall from now 
on omit to write any trace over the representation ${\bf R}$ of the gauge group, since the way in which 
these traces appear can be reconstructed in an unambiguous way at any stage of the derivation.

\subsubsection{K\"ahler moduli space}

The effective K\"ahler potential for the K\"ahler moduli, ignoring matter fields, is well known \cite{DKL,CdlO}. 
It can be derived in a quite straightforward way by retaining only the terms depending quadratically on 
space-time derivatives of the fields $G_{i \bar \jmath}$ and $B_{i \bar \jmath}$ in (\ref{kinterms}). To work 
out the reduction, one considers the real $(1,1)$ forms $J$ and $B$ associated to these two fields and 
decomposes the complex combination $J + i B$ onto the basis of real harmonic $(1,1)$ forms $\omega_A$,
with complex coefficients $T^A$ defining the 4D complex moduli fields: 
\be
J + i B = 2\,T^A \omega_A \,. 
\ee
In components this means $G_{i \bar \jmath} = -i(T^A + \bar T^A) \omega_{A i \bar \jmath}$ 
and $B_{i \bar \jmath} = -i(T^A - \bar T^A) \omega_{A i \bar \jmath}$. Plugging these decompositions 
into the first two terms of (\ref{kinterms}), one then finds a kinetic term for the complex scalar fields 
$T^A$ of the form
\be
{\cal L}_4 = - g_{A \bar B}^{\rm mod} \, \partial_\mu T^A \partial^\mu \bar T^B\,,
\ee
where 
\bea
g_{A \bar B}^{\rm mod} \b=\b - \frac 1{V} \int \! d^6 y \, \sqrt{G} \, G^{i \bar \jmath} G^{p \bar q} 
\omega_{A i \bar q} \, \omega_{B p \bar \jmath} \nn \\
\b=\b \frac 1{V} \int_X \!\! \omega_A \wedge *\omega_B \,.
\label{modmetric}
\eea
This metric does not depend at all on the forms $c_{PQ}$, and the issue of whether these 
are harmonic or not is therefore trivially irrelevant here. Using the decomposition 
$J = J^A \omega_A$ with $J^A = T^A + \bar T^A$, which implies that $\partial_A J^B = \delta_A^B$, 
and the relation (\ref{rel3}), one can rewrite (\ref{modmetric}) in the following form:
\bea
g_{A \bar B}^{\rm mod} \b=\b \frac 1{V} \int_X \!\! \omega_A \wedge *\omega_B \nn \\[1mm]
\b=\b - \partial_A \partial_{\bar B} \log V \,.
\eea
From this expression we deduce that the K\"ahler potential is given, up to a K\"ahler transformation, by:
\be
K = - \log V \,.
\ee
This can finally be rewritten more explicitly in terms of the chiral multiplets $T^A$ and the intersection numbers 
$d_{ABC}$ as 
\be
K = - \log \Big[\frac 16 d_{ABC} J^A J^B J^C\Big] \,,\;\; \text{with} \;\; J^A = T^A + \bar T^A \,.
\label{KT}
\ee
This result has the special property of being special-K\"ahler and also of the no-scale type, with the property:
\be
K_A K^A = 3 \,.
\label{noscaleT}
\ee
Notice finally that in geometrical terms the quantities $K_A$ and $K^A$ have 
the following simple expressions:
\be
K_A = - \frac 1{V} \int_X \!\! \omega_A \wedge *J \,,\;\;
K^{A} = - \int_X \!\! \omega^A \wedge J \,. \label{relKA}
\ee

\subsubsection{Matter field metric}\label{Mattermetric}

Let us next consider the addition of matter fields, under the simplifying assumption that their background 
value vanishes or is very small. In this situation, all the terms involving the fields $A_i$ without 
space-time derivatives can be neglected in (\ref{kinterms}), and the only term to be added is therefore the last one.
In this limit the matter sector can be considered as a small perturbation to the moduli sector, and one 
can neglect the interference between these two sectors. To work out the reduction, one may consider 
the $1$-forms $A$ taking values in the representation $({\bf R},{\bf r)}$ of $G \times S$, 
and decompose them onto the basis of harmonic $1$-forms $u_P$ taking values in the representation 
${\bf r}$ of $S$ with complex coefficients $\Phi^P$ taking values in the representation ${\bf R}$ of $G$ and 
defining the 4D matter fields:
\be
A = \Phi^P u_P \,.
\ee
In components this means $A_i = \Phi^P u_{Pi}$. Plugging this decomposition
into the last term of (\ref{kinterms}), one then finds a kinetic term for the complex 
scalar fields $\Phi^P$ of the form
\be
{\cal L}_4 = - g_{P \bar Q}^{\rm mat} \partial_\mu \Phi^P \partial^\mu \bar \Phi^Q \,,
\ee
where
\bea
g_{P \bar Q}^{\rm mat} \b=\b - \frac i{V} \int \! d^6y \, \sqrt{G} \, G^{i \bar \jmath} \, c_{PQ i \bar \jmath} \nn \\
\b=\b \frac 1{V} \int_X \!\! c_{PQ} \wedge *J \,.
\label{gphigen}
\eea
This metric depends on the forms $c_{PQ}$, but only through their scalar product with the K\"ahler 
form $J$, which is harmonic. As a result, only the harmonic component of the Hodge decomposition of 
$c_{PQ}$ matters, and the issue of whether the whole forms $c_{PQ}$ are harmonic or not is 
therefore again irrelevant. Using the decomposition $J = J^A \omega_A$ with $J^A = T^A + \bar T^A$, 
which as before implies that $\partial_A J^B = \delta_A^B$, as well as the decomposition of $*J$ on the 
dual basis $\omega^A$ and the relation (\ref{rel2}), one may rewrite (\ref{gphigen}) in the following form:
\bea
g_{P \bar Q}^{\rm mat} \b=\b \frac 1{V} \int_X \!\! \omega_A \wedge *J 
\int_X \!\! c_{PQ} \wedge \omega^A \nn \\[1mm]
\b=\b \partial_A \log V c^A_{PQ} \,.
\eea
This means that the matter metric is linked to the moduli K\"ahler potential by the relation
$g_{P \bar Q}^{\rm mat} = - K_A c^A_{PQ}$ \cite{BO,PS}. This in turn implies that the leading 
matter-dependent correction to the K\"ahler potential is given by this metric contracted with 
two matter fields. It should however be emphasized that this not only reproduces the 
matter kinetic terms analyzed in this subsection, but also induces a kinetic mixing between 
matter and moduli fields proportional to one matter field, as well as a correction to the 
moduli metric proportional to two matter fields. These terms do indeed occur, as will be 
clarified in next subsection, but they are negligible under the assumptions made here, and 
the leading correction to the K\"ahler potential is indeed
\be
\Delta K = - K_A c^A_{PQ} \Phi^P \bar \Phi^Q\,.
\label{DKT}
\ee
Notice finally that one can write the following simple geometric expressions for the contractions 
$K_A c^A_{PQ}$ and $K_{AB} c^B_{PQ}$:
\bea
\a\a K_A c^A_{PQ} = - \frac 1{V} \int_X \!\! c_{PQ} \wedge *J \,, \label{contr1} \\
\a\a K_{AB} c^B_{PQ} = \frac 1{V} \int_X \!\! \omega_A \wedge *c_{PQ} \label{contr2} \,.
\label{KABl}
\eea

\subsubsection{Full scalar manifold}

Let us finally consider the full dependence on both the K\"ahler moduli and the matter fields, which 
is relevant when the matter fields have a non-vanishing and sizable VEV. In this case, one has to 
consider all the terms in (\ref{kinterms}). The relevant fields are as before $G_{i \bar \jmath}$, 
$B_{i \bar \jmath}$ and $A_i$. The first two can be combined to form a complex $(1,1)$ form $J + i B$, 
and decomposed onto the basis of harmonic  $(1,1)$ forms $\omega_A$. The second can be viewed 
as matrix-valued $1$-forms $A$, and decomposed onto the basis of harmonic $1$-forms $u_P$. 
It however turns out that that the precise definition of the 4D moduli fields $T^A$ and matter fields 
$\Phi^S$ that allows to recast the action in a manifestly supersymmetric form involves a 
non-trivial shift. The form of this shift may be guessed by generalizing the results applying in the 
two special cases of Calabi-Yau manifolds with a single modulus and of orbifolds, which are also 
the only two cases where a derivation of the full effective K\"ahler potential is already known, 
respectively from \cite{W} and \cite{Korb1}. The only quantity that can possibly enter in the non-trivial 
shift is $c^A_{PQ}$, and the appropriate definitions turn out to be
\be
J + i B = 2 \Big(T^A - \frac 12 c^A_{PQ} \Phi^P \bar \Phi^Q \Big) \omega_A \,,\;\;
A = \Phi^P u_P \,.
\label{decompgen}
\ee
In components this means $G_{i \bar \jmath} = -i(T^A + \bar T^A - c^A_{PQ} \Phi^P \bar \Phi^Q) \omega_{A i \bar \jmath}$,
$B_{i \bar \jmath} = -i(T^A - \bar T^A) \omega_{A i \bar \jmath}$ and $A_i = \Phi^P u_{Pi}$. 
Plugging these decompositions into (\ref{kinterms}), one then finds kinetic terms for the complex scalar fields 
$T^A$ and $\Phi^P$ of the form
\be
{\cal L}_4 = - g_{A \bar B}^{\rm mod} \, \partial_\mu T^A \partial^\mu \bar T^B
- g_{P \bar Q}^{\rm mat} \, \partial_\mu \Phi^P \partial^\mu \bar \Phi^Q
- \big(g_{A \bar Q}^{\rm mix} \, \partial_\mu T^A \partial^\mu \bar \Phi^Q 
+ {\rm c.c.} \big)\,,
\ee
where 
\bea
g_{A \bar B}^{\rm mod} \b=\b - \frac 1{V} \int \! d^6 y \, \sqrt{G} \, G^{i \bar \jmath} G^{p \bar q} \, \omega_{A i \bar q} \, \omega_{B p \bar \jmath} \nn \\
\b=\b \frac 1{V} \int_X \!\! \omega_A \wedge *\omega_B \,, \label{metmod}
\eea
\vspace{-20pt}
\bea
\hspace{1.5pt} g_{P \bar Q}^{\rm mat} \b=\b - \frac i{V} \int \! d^6y \, \sqrt{G} \, G^{i \bar \jmath} \, c_{PQ i \bar \jmath} 
- \frac 1{V} \int \! d^6 y \, \sqrt{G} \, G^{i \bar \jmath} G^{m \bar n} \, c_{P S i \bar n} \, c_{R Q m \bar \jmath} \,
\Phi^R \bar \Phi^{S} \nn \\
\b=\b \frac 1{V} \int_X \!\! c_{PQ} \wedge *J 
+ \frac 1{V} \bigg\{\int_X \!\! c_{PS}  \wedge *c_{RQ} \bigg\} \Phi^R \bar \Phi^{S}\,, \label{metmat}
\eea
\vspace{-20pt}
\bea
\hspace{2.5pt} g_{A \bar Q}^{\rm mix} \b=\b 
\frac 1{V} \int \! d^6 y \, \sqrt{G} \, G^{i \bar \jmath} G^{m \bar n} \, \omega_{A i \bar n} \, c_{RQ m \bar \jmath} \,\Phi^R \nn \\
\b=\b - \frac 1{V} \bigg\{\int_X \!\! \omega_A \wedge *c_{RQ} \bigg\} \Phi^R \,. \label{metmix}
\eea
This metric now significantly depends on the forms $c_{PQ}$, not only through their scalar product with the K\"ahler 
form $J$ or the basis forms $\omega_A$, which are harmonic, but also through their scalar products among themselves.
As a result, not only the harmonic part but also the exact and coexact parts of the Hodge decomposition of $c_{PQ}$
matter, and the issue of whether $c_{PQ}$ is harmonic or not is therefore crucial in this case. As already said, we shall
for the moment assume that $c_{PQ}$ is harmonic and $c^A_{PQ}$ is constant, so that one can use the 
decomposition $c_{PQ} = c^A_{PQ} \omega_A$. Taking into account the new decomposition 
$J = J^A \omega_A$ with $J^A = T^A + \bar T^A - c^A_{PQ} \Phi^P \bar \Phi^Q$, which still implies that 
$\partial_A J^B = \delta_A^B$ since $c^A_{PQ}$ is constant, and using the relations (\ref{rel2}) and (\ref{rel3}), 
the metric components (\ref{metmod}), (\ref{metmat}) and(\ref{metmix}) can be rewritten as 
\bea
g_{A \bar B}^{\rm mod} \b=\b \frac 1{V} \int_X \!\! \omega_A \wedge *\omega_B \nn \\[2mm]
\b=\b - \partial_A \partial_{\bar B} \log V \,, \label{metmod2}
\eea
\vspace{-22pt}
\bea
\hspace{1.5pt} g_{P \bar Q}^{\rm mat} \hspace{0.7pt} \b=\b \frac 1{V} \bigg\{\int_X \!\! \omega_A \wedge *J \bigg\} c^A_{PQ}
+ \frac 1{V} \bigg\{\int_X \!\! \omega_A \wedge *\omega_B \bigg\} c^A_{PS} c^B_{RQ} \Phi^R \bar \Phi^S \nn \\[1.5mm]
\b=\b \partial_A \log V c^A_{PQ} - \partial_A \partial_{\bar B} \log V 
c^A_{PS} c^B_{RQ} \Phi^R \bar \Phi^S \nn \\[3.5mm]
\b=\b - \partial_P \partial_{\bar Q} \log V \,, \label{metmat2}
\eea
\vspace{-22pt}
\bea
\hspace{2.5pt} g_{A \bar Q}^{\rm mix} \b=\b - \frac 1{V} \bigg\{\int_X \!\! \omega_A \wedge *\omega_B \bigg\} c^B_{RQ} \Phi^R \nn \\[2mm]
\b=\b \partial_A \partial_{\bar B} \log V c^B_{RQ} \Phi^R \nn \\[3.5mm]
\b=\b - \partial_A \partial_{\bar Q} \log V \,. \label{metmix2}
\eea
From these expressions we see that, modulo an arbitrary K\"ahler transformation, the K\"ahler potential is simply given by:
\be
K = - \log V \,.
\ee
More explicitly, this reads in this case:
\be
K = - \log \Big[\frac 16 d_{ABC} J^A J^B J^C\Big] \,,\;\; \text{with} \;\; J^A = T^A + \bar T^A - c^A_{PQ} \Phi^P \bar \Phi^Q \,.
\label{Kgen}
\ee
This result coincides with the one proposed in \cite{PS} on the basis of an $M$-theory argumentation. It manifestly 
reproduces the result (\ref{KT}) for the moduli and the leading order correction (\ref{DKT}) at quadratic order in the 
matter fields. Moreover its satisfies a no-scale property generalizing (\ref{noscaleT}). This is easily demonstrated 
as follows \cite{PS}. Since $e^{-K}$ is homogenous of degree $3$ in $J^A$, we have $J^A \partial K/\partial J^A = - 3$. 
Denoting the fields by $Z^I = T^A,\Phi^P$, we then compute $K_I = \partial K/\partial J^A \partial J^A/\partial Z^I$. 
In particular, $K_A = \partial K/\partial J^A$ so that $K_A J^A = - 3$. Taking a derivative of this relation with respect to 
$\bar Z^{\bar J}$, it follows that $K_{\bar J A} J^A + K_A \partial J^A/\partial \bar Z^{\bar J}  = 0$, or 
$K_{\bar J A} J^A + K_{\bar J} = 0$. Finally, acting on this with the inverse metric $K^{I \bar J}$ one deduces that 
$K^I = - \delta^I_A J^A$. It finally follows that $K_I K^I = - K_A J^A = 3$. Splitting again the two kinds of indices,
this means:
\be
K_A K^A + K_P K^P = 3 \,.
\label{noscalegen}
\ee
Notice finally that $K_A$, $K_P$, $K^A$ and $K^P$ can be written in the following simple geometrical
terms:
\bea
\a\a K_A = - \frac 1{V} \int_X \!\! \omega_A \wedge *J \,,\;\;
K^{A} = - \int_X \!\! \omega^A \wedge J \,, \label{relKAnew} \\
\a\a K_P = \frac 1{V} \int_X \!\! c_{PS} \bar \Phi^S \wedge *J \,,\;\; 
K^P = 0 \,. \label{relKSnew}
\eea
Moreover, from the assumption that the forms $c_{PQ}$ are harmonic it follows that
also the contraction $K_{A B} c^A_{PQ} c^B_{RS}$ admits a simple geometrical
expression:
\be
K_{A B} c^A_{PQ} c^B_{RS} = \frac 1{V} \int_X c_{PQ} \wedge *c_{RS} 
\label{contr3}\,.
\ee
Similarly one also finds that
\be
d_{A B C} c^A_{PQ} c^B_{RS} c^C_{MN} = 
\int_X c_{PQ} \wedge c_{RS} \wedge c_{MN}
\label{contr4}\,.
\ee

\subsection{Range of validity}

The simple derivation presented in last subsection is manifestly valid in those cases where the 
forms $c_{PQ}$ are harmonic and the quantities $c^A_{PQ}$ are constant topological invariants. 
One special situation in which this is certainly true is when all the involved forms $\omega_A$ and $u_P$
are actually not only harmonic but actually covariantly constant. As we shall see more explicitly in 
next section, this is for instance the case for toroidal orbifold models. But we believe that it could 
be true also in a less trivial fashion. We will imagine that this is indeed the case for some subset 
of smooth Calabi-Yau models. For further use, let us then explore a few simple consequences of the 
above assumptions. Recall that $A=0,\cdots,h^{1,1}-1$ labels the different K\"ahler moduli and 
$P,Q=1,\cdots,n_R$ label the different matter fields. By definition, for each of the $h^{1,1}$ values 
of $A$ the quantity $c^A_{PQ}$ is a Hermitian $n_R \times n_R$ matrix. This means that even when 
$h^{1,1} > n_R^2$, the number of these matrices that are linearly independent can not exceed 
$n_R^2$. In fact, the $h^{1,1}$ matrices $c^A_{PQ}$ can always be rewritten as linear 
combinations of the $n_R^2$ independent transposed Hermitian matrices $\lambda^{A'}_{QP}$, 
with $A'=0,\cdots,n_R^2-1$ and where the transposition is included for later convenience. Notice 
that whereas the matrices $c^A_{PQ}$ do a priori not satisfy any completeness relation and do 
not generate any closed algebra, the matrices $\lambda^{A'}_{PQ}$ do instead satisfy an obvious 
completeness relation since they form a basis of Hermitian matrices and generate a closed algebra, 
which is that of $U(n_R)$. We therefore know that under the assumptions that we made
\bea
\a\a c^A_{PQ} \;:\;\; \text{linear combinations of}\; \lambda^{A'}_{QP} \,,\\
\a\a \,\lambda^{A'}_{PQ}\,\,:\;\; \text{$n_R \times n_R$ matrices representing the generators of $U(n_R)$} \,.
\eea

The extension to more general situations where instead $c_{PQ}$ is not harmonic 
and the quantities $c^A_{PQ}$ are not constant topological invariants is clearly more 
challenging, and one may wonder whether a result similar to (\ref{Kgen}) could hold true. 
One first major change arising for a non-harmonic $c_{PQ}$ is that since its Hodge 
decomposition contains now not only a harmonic piece but also an exact piece and 
a coexact piece, eq.~(\ref{contr3}) does no longer hold true. More precisely, its left-hand 
side acquires extra terms matching the contributions to the right-hand side coming from 
the non-harmonic parts of $c_{PQ}$, which are clearly more difficult to deal with. In 
particular, when going from (\ref{metmat}) to (\ref{metmat2}), one would get additional 
terms that clearly have to do with the effect of heavy non-zero modes. In fact, these 
heavy modes must be related to the 10D $B$ field. Indeed, using a democratic formulation 
of the original 10D theory involving not only the $2$ form $B$ but also its magnetic dual 
$6$ form $\tilde B$, the contact term from which the problem originates can be 
deconstructed and the seed for its origin is then reduced to a linear coupling between 
$\tilde B$ and $d \Gamma = {\rm tr} (F \wedge F)$. When reducing on $X$, one then gets 
a direct coupling between two light matter modes coming from $A$ and one heavy mode 
coming from $\tilde B$ whenever $c_{PQ}$ is not harmonic, and this must be responsible 
form the extra contributions to the contact terms. 
A second source of difficulty arising for a non-constant $c^A_{PQ}$ is that this quantity may 
then be expected to depend on continuous deformations of both the vector bundle $V$ and 
the manifold $X$. The first of these dependences, which was already mentioned in \cite{PS}, 
does not concern us since it would be related to vector bundle moduli, which we have ignored 
from the beginning. But the second of these dependences, which we believe should also be a 
priori feared, is instead directly relevant for our derivation, since it is related to the K\"ahler 
moduli that we want to keep in the effective theory. Now, a moduli dependence $c^A_{PQ}$
would imply additional terms in (\ref{metmod})--(\ref{metmix}). Moreover,
it would also affect the simple relation $\partial_A J^B = \delta_A^B$ that was used to rewrite 
these metric in the form (\ref{metmod2})--(\ref{metmix2}). At first one might hope 
that these two sources of complications could compensate each other, but things do not 
seem to be so simple. One may then perhaps have to generalize the decomposition 
(\ref{decompgen}) through a more complicated and implicit definition of the moduli and matter fields. 
We were however not able to reach a conclusive assessment of this possibility.

We believe that subtleties very similar to those explained here for heterotic models may actually 
arise also for orientifold models. More precisely, it seems to us that the results derived in 
\cite{Korient2,Korient3} concerning the higher-order dependence of the K\"ahler potential on the matter 
fields arising from $D$-brane sectors should a priori also be correct and reliable only for 
those special models were massive non-zero modes do not induce non-trivial corrections.
We attribute the fact that this is not directly signaled by a technical difficulty in the derivation 
of \cite{Korient2,Korient3} to the use of a democratic formulation in terms of all the 
Ramond-Ramond forms, which deconstructs the original 10D contact term and hides the subtlety.

\subsection{Standard embedding}

The concerns raised in previous subsection may be illustrated more concretely by considering in some detail the 
special case of Calabi-Yau manifolds $X$ with a generic number of moduli but standard embedding for the vector bundle $V$. 
In this case the situation is somewhat simpler and there exist an alternative way of performing the dimensional 
reduction for the matter fields. Indeed, recall that in this case $V$ is identified with $TX$, so that $S=SU(3)$ and 
$G = E_6 \times E_8$. As a consequence, the additional index in the representation ${\bf r} = {\bf \bar 3}$ can 
be reinterpreted as a cotangent space index, and one may exploit this to construct the $SU(3)$-valued 
harmonic $1$ forms $u_A$ in terms of the harmonic $(1,1)$ forms $\omega_A$. 

In the approximation where one works at leading order in the matter fields and neglects the interference 
between moduli and matter fields, as in subsection \ref{Mattermetric}, the way in which this decomposition 
can be done has been explained in \cite{GLM} (see also \cite{BLM}). In the end, it essentially amounts 
to describe the matter modes in terms of a standard $(1,1)$ form $\tilde A$ and decompose it on the basis of 
harmonic $(1,1)$ forms $\omega_A$ with $h^{1,1}$ complex coefficients $\Phi^A$ taking values in the 
representation ${\bf R} = ({\bf \overline{27}}, {\bf 1})$ of $E_6 \times E_8$ and defining the 4D matter fields. 
It has been shown in \cite{GLM} that one must however include a suitable power of the norm of the covariantly 
constant holomorphic $(3,0)$ form of $X$ in this decomposition, in order to be able to express the potential coming 
from the non-derivative part of the action in terms of a holomorphic superpotential. Here, since we are 
considering the case of absent or frozen complex structure moduli, this simply implies some extra power 
of the volume $V$, and the correct definition turns out to be 
\be
\tilde A = V^{1/6} \Phi^A \omega_A \,.
\label{Astand}
\ee
One then finds a kinetic term of the form
\be
{\cal L}_4 = - g_{A \bar B}^{\rm mat} \partial_\mu \Phi^A \partial^\mu \bar \Phi^B \,,
\ee
where
\bea
g_{A \bar B}^{\rm mat} \b=\b - \frac 1{V^{2/3}} \int \! d^6 y \, \sqrt{G} \, G^{i \bar \jmath} G^{p \bar q} \, 
\omega_{A i \bar q} \, \omega_{B p \bar \jmath} \nn \\
\b=\b \frac 1{V^{2/3}} \int_X \!\! \omega_A \wedge *\omega_B \,.
\label{gphistand}
\eea
Through the usual manipulations, this metric can be rewritten as
\be
g_{A \bar B}^{\rm mat} = - V^{1/3} \partial_A \partial_{\bar B} \log V \,.
\ee
This implies that the matter metric is in this case linked to the moduli metric by the
relation $g_{A \bar B}^{\rm mat} = e^{- K/3} g_{A \bar B}^{\rm mod}$, which was first derived 
in \cite{DKL} by matching an actual string scattering amplitude computation. The leading 
matter-dependent correction to the moduli K\"ahler potential must then have the form
\be
\Delta K = e^{- K/3} K_{A \bar B} \Phi^A \bar \Phi^B \,. \label{DKTstand}
\ee
Comparing the result (\ref{DKTstand}) with the general expression (\ref{DKT}) and requiring 
them to be equal, we deduce that in the case of standard embedding the matrices $c^A_{BC}$ 
must have a special form. This is indeed the case. The components of the $(1,1)$ form 
$c_{AB}$ are found to be given by
\be
c_{AB i \bar \jmath} = - i\,V^{1/3} G^{p \bar q} \, \omega_{A i \bar q} \, \omega_{B p \bar \jmath} \,.
\label{lambdacomp}
\ee
It is a straightforward exercise to verify that the forms $c_{AB}$ defined by these components are
generically not harmonic, except for the particular case where $\omega_A$ and/or $\omega_B$ 
is identified with the K\"ahler form $J$ or happen more in general to be a covariantly constant 
$(1,1)$ form. Since by eq.~(\ref{relKA}) one has $K^A \omega_A = - J$, this means that:
\be
c_{AB} \;\,\text{not harmonic}\,,\;\; \text{but}\;\, K^A c_{AB} \;\,\text{and}\;\, K^B c_{AB} \;\, \text{harmonic} \,. 
\ee
One may nevertheless compute the quantity $c^A_{BC}$ by using the expression (\ref{lambdacomp}) for the 
components of $c_{PQ}$. The result depends on the metric and is thus a function of $T^D + \bar T^D$. It might 
be possible to express this function in terms of derivatives of the K\"ahler potential $K$ for the moduli. But 
even without writing an explicit expression, one can observe that the factor $V^{1/3} G^{p \bar q}$ appearing in the 
expression (\ref{lambdacomp}) is a homogenous function of degree $0$ in the components of the metric, and 
therefore in the geometric moduli fields. More precisely, one finds that $c^0_{00} = 1$ when $h^{1,1} = 1$ and 
there is a single modulus $T^0$, whereas $c^A_{BC} = c^A_{BC}((T^D + \bar T^D)/(T^E + \bar T^E))$ 
when $h^{1,1} > 1$ and there are several moduli $T^A$. Since by eq.~(\ref{relKA}) one has $K^D = -(T^D + \bar T^D)$, 
this means that
\be
c^A_{BC} \;\, \text{not constant}\,,\;\;\text{but}\;\, K^D \partial_D c^A_{BC} = 0 \,.
\label{hom}
\ee
Finally, using the relation (\ref{contr1}) and the expression (\ref{lambdacomp}), one easily verifies that 
$c^A_{BC}$ does indeed satisfy an identity ensuring that the two expressions (\ref{DKT}) and (\ref{DKTstand}) 
are identical:
\be
- K_A c^A_{BC} = e^{-K/3} K_{B C} \,.
\label{relstand}
\ee
One can demonstrate analytically that the above relation forces $c^A_{BC}$ to be constant in the special 
case $h^{1,1} = 1$ and non-constant when instead $h^{1,1} > 0$. To do so, one starts by assuming that 
(\ref{relstand}) is satisfied with a constant $c^A_{BC}$. One may then take a derivative of (\ref{relstand}), 
use $\partial_D c^A_{BC} = 0$ and act with the inverse of the moduli metric to derive the expression 
$c^{A}_{BC} = - e^{-K/3} K^{A D} \big(K_{B C D} - \frac 13 K_{D} K_{B C}\big)$. Finally, one may compute the 
derivative of this expression to check whether it is really zero, as assumed. In particular, using the identity 
$\partial_A K^B = -\delta_A^B$ one finds rather easily that $\partial_A c^A_{BC} K^B K^C = - 3 \, e^{- K/3} \big(h^{1,1} - 1 \big)$, 
which vanishes when $h^{1,1} = 1$ but not when $h^{1,1} > 1$, contradicting in this last case the hypothesis
that $c^A_{BC}$ was constant.

When attempting to go on and work out the result at higher orders in the matter fields, one can no longer neglect 
the interference between matter and moduli fields. One then needs to properly change the definition of the moduli 
fields. The natural guess based on our general derivation is that the definition of the moduli fields should be shifted 
by a term that is quadratic in the matter fields and involves $c^A_{BC}$. Indirect evidence in favor of this has 
been found in \cite{GLM} (whose quantity $\sigma_{ABC}$ is seen to be proportional to our $c^A_{BC}$ 
specified by (\ref{lambdacomp}) with the upper index lowered with the moduli metric) by studying the interference 
of this redefinition and the possible emergence of a non-trivial superpotential. It is however not obvious how one 
should proceed to work out the full result, as both of the subtleties discussed in section 2.3, namely the non-harmonicity
of $c_{BC}$ and the non-constancy of $c^A_{BC}$, have been manifestly shown to arise in this case, except for the 
particular situations where $h^{1,1} = 1$, for which the result (\ref{Kgen}) holds true and reduces to the result derived 
in \cite{W}.

\section{The heterotic string on an orbifold}
\setcounter{equation}{0}

It is interesting to compare the general situation occurring for compactifications on a smooth Calabi-Yau manifold $X$
with that of compactifications on toroidal orbifolds of the type $T^6/Z_N$ \cite{Torhet}, which represent singular limits of them from 
the geometrical point of view. We shall briefly review the structure of these models and the derivation of the effective
K\"ahler potential. We shall as before focus on the K\"ahler moduli and the matter fields, restricting to the untwisted sector 
for which a simple derivation based on dimensional reduction is possible, and show how the known exact results 
for the dependence of the K\"ahler potential on the K\"ahler moduli and matter fields can be rephrased in the same 
language as in the previous section.

\subsection{General structure}

The $Z_N$ orbifold action that is used to define the background is specified by a first twist vector 
$\alpha_i = (\alpha_1,\alpha_2,\alpha_3)$ in the $SU(3)$ internal space-time group and a second twist vector 
$\beta_\alpha = (\beta_1, \cdots,\beta_8;\beta_1,\cdots,\beta_8)$ in the $E_8\times E_8$ gauge group. 
These should satisfy the following consistency condition for some integer $n$, which comes from the level-matching 
condition \cite{Torhet}:
\be
\frac nN = \sum_i \alpha_i (\alpha_i + 1) - \sum_\alpha \beta_\alpha (\beta_\alpha + 1) \,. \label{levelmatch}
\ee
From a geometric perspective, the choice of $\alpha_i$ defines the structure group of the tangent space to be 
a discrete subgroup of $SU(3)$, whereas the choice of $\beta_\alpha$ corresponds to a choice of vector bundle. 
The condition (\ref{levelmatch}) is the analogue of the Bianchi identity (\ref{bianchi}) that must be imposed for
smooth Calabi-Yau compactifications and constrains the choice of vector bundle for a given tangent bundle. This 
leaves as before several possibilities, among which one again finds the special possibility of the standard 
embedding, which corresponds to the choice $\beta_\alpha = (\alpha_1,\alpha_2,\alpha_3, 0, \cdots, 0)$ 
and trivially satisfies (\ref{levelmatch}) with $n=0$. 

The states arising in the untwisted sector are associated to the subset of harmonic forms on $T^6$ that are left 
invariant by the $Z_N$ twist. As a result, the low-energy effective theory can easily be computed and turns out 
to be a projection of what would be obtained by compactifying on $T^6$. The spectrum of neutral fields can be 
understood by looking at the transformation properties of the various harmonic forms under the discrete structure 
group $Z_N \subset SU(3)$ of $TX$. One in particular sees that the ${\bf 1}$ is always kept and the ${\bf 3}$ is 
always lost, whereas $h^{1,2}$ forms in the ${\bf 6}$ and $h^{1,1}-1$ forms in the ${\bf 8}$ survive the projection, 
with $h^{1,1}$ and $h^{1,2}$ being the effective Hodge numbers pertaining to the untwisted sector. We will restrict 
to the prototypical cases based on $N=3,6,7$, which lead to $h^{1,1} = 9,5,3$ and $h^{1,2} = 0$. The spectrum 
of charged fields can similarly be understood by looking at the transformation properties of the various forms not only 
under the discrete structure group $Z_N \subset SU(3)$ of $TX$, but also under the discrete structure group $S$ of $V$.

The simplest models are obtained by choosing bundles whose structure group is either trivial or a discrete $Z_N$ 
subgroup of $SU(3)$ in each of the two sectors. In the first case one has $E_8 \to E_8$ with ${\bf 248} \to {\bf 248}$, 
and the gauge group is unbroken. In the second case on has $E_8 \to E_6 \times SU(3)$ with 
${\bf 248} \to ({\bf 78},{\bf 1}) \oplus ({\bf 1},{\bf 8}) \oplus ({\bf 27},{\bf 3}) \oplus ({\bf \overline{27}},{\bf \overline 3})$,
and further $SU(3) \to H$ with ${\bf 3} \to {\bf h}$, so that the gauge group is broken to $E_6 \times H$, where the 
enhanced gauge symmetry $H \subset SU(3)$ arises as the non-trivial commutant of the discrete structure 
group $Z_N$ within $SU(3)$. In the three models under consideration, one respectively finds the three possible maximal-rank 
subgroups $H=SU(3), SU(2) \times U(1), U(1) \times U(1)$, with ${\bf h} = {\bf 3}, {\bf 2 \oplus 1}$, ${\bf 1} \oplus {\bf 1} \oplus {\bf 1}$. 
The various generations of untwisted matter fields in the ${\bf 27}$ and ${\bf \overline{27}}$ of $E_6$ must then arrange
into the representations ${\bf h}$ and ${\bf \bar h}$ of $H$ descending form the ${\bf 3}$ and ${\bf \bar 3}$ of $SU(3)$. 
In order to compare with the case of smooth Calabi-Yau manifolds and make it simpler, let us for a moment count the 
total numbers $n_{\rm 27}$ and $n_{\overline{27}}$ of ${\bf 27}$ and ${\bf \overline{27}}$ without caring about the 
$H$ quantum numbers. A first type of model can be constructed by making the asymmetric choice 
$S_{\rm v} = Z_N$, $S_{\rm h} = \text{trivial}$. One then finds $G_{\rm v} = E_6 \times H$ 
and $n_1^{\rm v} = 0$, $n_{27}^{\rm v} = 0$, $n_{\overline{27}}^{\rm v} = h^{1,1}$ in the visible sector, and just 
$G_{\rm h} = E_8$ in the hidden sector. The standard embedding is a particular case of this class of models 
where the level matching condition is trivially satisfied. A second type of model can be constructed by making the 
symmetric choice $S_{\rm v} = Z_N$, $S_{\rm h} = Z_N$. One then finds $G_{\rm v} = E_6 \times H$ and 
$n_1^{\rm v} = 0$, $n_{27}^{\rm v} = 0$, $n_{\overline{27}}^{\rm v} = h^{1,1}$  in the visible sector, and similarly 
$G_{\rm h} = E_6 \times H$ and $n_1^{\rm h} = 0$, $n_{27}^{\rm h} = 0$, $n_{\overline{27}}^{\rm h} = h^{1,1}$ 
in the hidden sector. In addition, there always are $h^{1,1}$ K\"ahler moduli.

\subsection{Effective K\"ahler potential}

The 4D effective K\"ahler potential for the untwisted sector of orbifold models is most easily computed by simply
retaining those fields that are invariant under the $Z_N$ projection in (\ref{kinterms}). One can then compute the metric,
guess the appropriate definition of the chiral multiplets that makes this manifestly K\"ahler, and finally find out 
the form of the K\"ahler potential. This last step can be done by relying on some basic properties of square matrices, 
which are described at the end of appendix A. Here we would like to emphasize that the same result can 
be obtained by proceeding exactly as we did in section 2 for compactifications on smooth Calabi-Yau manifolds.
We shall briefly summarize how this is done for the three different kind of models under consideration, in order 
to make contact with the results of \cite{ASmild}. As before, for notational simplicity we shall omit to write explicitly 
the traces over the representation ${\bf R}$ of the gauge group $G$. We also omit any detail about the trace over the 
representation ${\bf r}$ of the structure group $S$, since this is discrete. Finally, we shall here restrict 
for concreteness to the particular models discussed at the end of the previous subsection.

\subsubsection{Models with $H=SU(3)$}

Let us first consider the case of the $Z_3$ orbifold, where $H=SU(3)$. In this case, 
$h^{1,1} = 9$ and $n_{({\bf \bar 3}, {\bf \overline{27}})} = 3$, so that in total $n_{\bf \overline{27}} = 9$.  
There are $9$ harmonic $(1,1)$ forms $\omega_{ij}$ and $3$ $Z_3$-valued harmonic $1$-forms $u_{i}$, 
with $i=1,2,3$:
\bea
\a\a \omega_{ij} = i\, d z^i \wedge d \bar z^j \,, \\
\a\a u_{i} = dz^i \,.
\eea
The intersection numbers are found to be:
\be
d_{ijpqrs} = \epsilon_{ipr} \epsilon_{jqr} \,.
\label{dorb1}
\ee
The forms $c_{ij} = i\, u_{i} \wedge \bar u_{j}$ are found to be given by 
$c_{ij} = \omega_{ij}$, and their components on the basis $\omega_{mn}$ 
read
\be
c^{mn}_{ij} = \delta^m_i \delta^n_j \,.
\label{lambdaorb1}
\ee
The moduli fields $T^{ij}$ and the matter fields $\Phi^{i}$ are then defined by the following 
expansions:
\bea
\a\a J + i B = 2 \Big(T^{ij} - \frac 12 \Phi^{i} \bar \Phi^{j} \Big)\omega_{ij} \,, \\
\a\a A = \Phi^{i} u_{i} \,.
\eea
The K\"ahler potential is finally found to be given by \cite{Korb1,Korb2}:
\bea
K \b=\b - \log \bigg[\det \Big(T^{ij} + \bar T^{ij} - \Phi^{i} \bar \Phi^j \Big) \bigg]\,.
\eea

\subsubsection{Models with $H=SU(2) \times U(1)$}

Let us next consider the case of the $Z_6$ orbifold, where $H=SU(2) \times U(1)$. In this case, 
$h^{1,1} = 5$ and $n_{({\bf \bar 2}, {\bf \overline{27}})} = 2$, $n_{({\bf 1}, {\bf \overline{27}})} = 1$, 
so that in total $n_{\bf \overline{27}} = 5$.  There are $5$ harmonic $(1,1)$ forms 
$\omega_{\underline{i}\underline{j}}$, $\omega_{33}$ and $3$ $Z_6$-valued harmonic $1$-forms 
$u_{\underline{i}}$, $u_3$, with $\underline{i}=1,2$:
\bea
\a\a \omega_{\underline{i}\underline{j}} = i\, d z^{\underline{i}} \wedge d \bar z^{\underline{j}} \,,\;\; 
\omega_{33} = i\, dz^3 \wedge d \bar z^3 \,, \\
\a\a u_{\underline{i}} = dz^{\underline{i}} \,,\;\; u_3 = dz^3 \,.
\eea
The non-vanishing entries of the intersection numbers are:
\be
d_{\underline{i}\underline{j}\underline{p}\underline{q}33} = \epsilon_{\underline{i}\underline{p}3} 
\epsilon_{\underline{j}\underline{q}3} \,.
\label{dorb2}
\ee
The forms $c_{ij} = i\, u_{i} \wedge \bar u_{j}$ are easily computed and one 
finds $c_{\underline{i}\underline{j}} = \omega_{\underline{i}\underline{j}}$,
$c_{33} = \omega_{33}$, while the other vanish. The non-vanishing components of these 
forms on the basis $\omega_{mn}$ are
\be
c^{\underline{m}\underline{n}}_{\underline{i}\underline{j}} = \delta^{\underline{m}}_{\underline{i}} 
\delta^{\underline{n}}_{\underline{j}} 
\,,\;\; c^{33}_{33} = 1\,.
\label{lambdaorb2}
\ee
The moduli fields $T^{\underline{i}\underline{j}}$, $T^{33}$ and the matter fields $\Phi^{\underline{i}}$, 
$\Phi^3$ are then defined by the following expansions:
\bea
\a\a J + i B = 2 \Big(T^{\underline{i}\underline{j}} - \frac 12 \Phi^{\underline{i}} \bar \Phi^{\underline{j}} \Big)
\omega_{\underline{i}\underline{j}} + 2 \Big(T^{33} - \frac 12 \Phi^3 \bar \Phi^3 \Big)\omega_{33} \,, \\
\a\a A = \Phi^{\underline{i}} u_{\underline{i}} + \Phi^3 u_3 \,.
\eea
The K\"ahler potential is finally found to be given by \cite{Korb1,Korb2}:
\bea
K \b=\b - \log \bigg[\det \Big(T^{\underline{i}\underline{j}} + \bar T^{\underline{i}\underline{j}} 
- \Phi^{\underline{i}} \bar \Phi^{\underline{j}} \Big) 
\Big(T^{33} + \bar T^{33} - \Phi^3 \bar \Phi^3 \Big)\bigg]\,.
\eea

\subsubsection{Models with $H=U(1) \times U(1)$}

Let us finally consider the case of the $Z_7$ orbifold, where $H=U(1) \times U(1)$. 
In this case, $h^{1,1} = 1$ and $n_{({\bf 1}, {\bf \overline{27}})} = 3$, so that in total $n_{\bf \overline{27}} = 3$. 
There are $3$ harmonic $(1,1)$ forms $\omega_{11}$, $\omega_{22}$, $\omega_{33}$ and $3$ $Z_7$-valued 
harmonic $1$-forms $u_1$, $u_2$, $u_3$:
\bea
\a\a \omega_{11} = i\, dz^1 \wedge d \bar z^1 \,,\;\;
\omega_{22} = i\, dz^2 \wedge d \bar z^2 \,,\;\; 
\omega_{33} = i\, dz^3 \wedge d \bar z^3 \,, \\
\a\a u_1 = dz^1 \,,\;\; u_2 = dz^2 \,,\;\; u_3 = dz^3 \,.
\eea
The non-vanishing entries of the intersection numbers are found to be:
\be
d_{112233} = 1 \,.
\label{dorb3}
\ee
The forms $c_{ij} = i\, u_{i} \wedge \bar u_{j}$ are found to be given 
by $c_{11} = \omega_{11}$, $c_{22} = \omega_{22}$, $c_{33} = \omega_{33}$,
while the other vanish. The non-vanishing components of these $c_{ij}$ on the basis 
$\omega_{mn}$ read:
\be
c^{11}_{11} = 1 \,,\;\; 
c^{22}_{22} = 1 \,,\;\; 
c^{33}_{33} = 1\,.
\label{lambdaorb3}
\ee
The moduli fields $T^{11}$, $T^{22}$, $T^{33}$ and the matter fields $\Phi^1$, $\Phi^2$, $\Phi^3$ are 
then defined by the following expansions:
\bea
\a\a J + i B = 2 \Big(T^{11} - \frac 12 \Phi^1 \bar \Phi^1 \Big)\omega_{11} 
+ 2 \Big(T^{22} - \frac 12 \Phi^2 \bar \Phi^2 \Big)\omega_{22} 
+ 2 \Big(T^{33} - \frac 12 \Phi^3 \bar \Phi^3 \Big)\omega_{33} \,,\\
\a\a A = \Phi^1 u_1 + \Phi^2 u_2 + \Phi^3 u_3 \,.
\eea
The K\"ahler potential is finally found to be given by \cite{Korb1,Korb2}:
\bea
K \b=\b - \log \bigg[\Big(T^{11}\! + \bar T^{11} \! - \Phi^1 \bar \Phi^1 \Big)
\Big(T^{22} \! + \bar T^{22} \! - \Phi^2 \bar \Phi^2 \Big)
\Big(T^{33} \! + \bar T^{33} \! - \Phi^3 \bar \Phi^3 \Big)\bigg]\,.
\eea

\subsubsection{General structure} \label{GS}

The above results can be rewritten in a more convenient and unified way by performing a suitable change 
of basis for the harmonic $(1,1)$ forms \cite{ASmild}, which clarifies their similarity with the results derived
for Calabi-Yau compactifications. To perform this change of basis, we can proceed in 
parallel for all the three models considered above and introduce the $3 \times 3$ Hermitian matrices 
$\lambda^A$ representing the generators of $U(1) \times H$ and normalized in such a way that 
${\rm tr}(\lambda^A \lambda^B) = \delta^{AB}$. More precisely, $\lambda^0$ denotes the generator of $U(1)$ proportional 
to the identity matrix and $\lambda^a$ the generators of $H$ associated to a subset of the Gell-Mann matrices 
spanning the fundamental representation of $SU(3)$ ($a=1,\cdots,8$ for $H=SU(3)$, 
$a=1,2,3,8$ for $H=SU(2)\times U(1)$, $a=3,8$ for $H=U(1) \times U(1)$): 
\be
\lambda^A_{ij} \,:\;\; \text{$3 \times 3$ matrices representing the generators of}\; U(1) \times H \,.
\ee 
We then define the new basis of harmonic $(1,1)$ forms $\omega_A = \lambda^A_{ij} \omega_{ij}$. The 
corresponding new moduli fields then read $T^A = \lambda^A_{ji} T^{ij}$, and since the matrices 
$\lambda^A$ are Hermitian, one finds $\bar T^A = \lambda^A_{ji} \bar T^{ij}$, where $\bar T^{ij}$ 
denotes as in the previous formulae the Hermitian conjugate of $T^{ij}$ as a matrix.
In this new basis, the intersection numbers are given by $d_{ABC} = \lambda^A_{ij} \lambda^B_{pq} \lambda^C_{rs} d_{ijpqrs}$, 
which yields
\be
d_{ABC} = 2\, {\rm tr} \big(\lambda^{(A} \lambda^B \lambda^{C)} \big) 
- 3\, {\rm tr} \big(\lambda^{(A}) {\rm tr} \big(\lambda^B \lambda^{C)} \big) 
+ {\rm tr} \big(\lambda^{(A}) {\rm tr} \big(\lambda^B \big) {\rm tr} \big(\lambda^{C)} \big) \,.
\label{dorb}
\ee
The components $c^A_{ij}$ of $c_{ij}$ are instead given by $c^A_{ij} = \lambda^A_{nm} c^{mn}_{ij}$, 
which simply gives:
\be
c^A_{ij} = \lambda^A_{ji} \,.
\label{lambdaorb}
\ee
In this basis, the fields are defined as 
\be
J + i B = 2 \Big(T^A - \frac 12 c^A_{ij} \Phi^{i} \bar \Phi^{j} \Big)\omega_A \,,\;\;
A = \Phi^{i} u_{i} \,,
\ee
and the K\"ahler potential takes the form:
\be
K = - \log \bigg[\frac 16 d_{ABC} J^A J^B J^C \bigg] \,,\;\; J^A = T^A + \bar T^A - c^A_{ij} \Phi^i \bar \Phi^j \,.
\ee
For the untwisted sector of these orbifolds, one thus finds exactly the same kind of result as for smooth Calabi-Yau 
manifolds, with the peculiarity, however, that the intersection numbers $d_{ABC}$ and the quantities $c^A_{ij}$ 
admit a group-theoretical interpretation. This corresponds to the fact that the scalar manifold becomes a symmetric 
space. More precisely, in the three kinds of models under consideration the scalar manifolds are given by:
\bea
\a\a {\cal M}_{SU(3)} = \frac {SU(3,3 + n)}{U(1) \times SU(3) \times SU(3 + n)} \,, \label{M1}\\
\a\a {\cal M}_{SU(2) \times U(1)} = \frac {SU(2,2 + n)}{U(1) \times SU(2) \!\times\! SU(2 + n)} 
\times \frac {SU(1,1 + n)}{U(1) \times SU(1 + n)}\,, \label{M2} \\
\a\a {\cal M}_{U(1) \times U(1)} = \frac {SU(1,1 + n)}{U(1) \times SU(1 + n)} \times
\frac {SU(1,1 + n)}{U(1) \times SU(1 + n)} \times 
\frac {SU(1,1 + n)}{U(1) \times SU(1 + n)} \,. \label{M3}
\eea

\subsection{Range of validity}

For the untwisted sector of orbifold models, we see that the low-energy effective K\"ahler potential can always be 
derived in an exact way, without any limitation. From the perspective of the more general study that we performed for 
smooth Calabi-Yau manifolds, this reflects the fact that untwisted orbifold sectors automatically satisfy the assumptions 
that we made in section 2. More specifically, we see that the forms $c_{ij}$ are harmonic and the quantities 
$c^A_{ij}$ are constants. This can be traced back to the fact that in this case the forms $\omega_A$ 
and $u_i$ are not only harmonic, but actually covariantly constant, which is a much stronger property.

\section{$M$-theory interpretation}
\setcounter{equation}{0}

The structure of the K\"ahler potential characterizing the 4D low-energy effective theories of heterotic string 
models admits a simple interpretation in terms of a 5D effective theory compactified on a segment $S^1/Z_2$, 
which describes the $M$-theory lift of these models. In particular, the definition of the chiral multiplets and the 
structure of the K\"ahler potential can be understood quite naturally and intuitively within this framework. 
As we shall briefly review in this section, this is a consequence of the fact that the matter contact terms arising 
from the non-trivial shift in the field-strength of the $2$ form $B$ in the heterotic picture arises in the $M$-theory 
picture from the exchange of the heavy Kaluza-Klein modes of the $3$ form $C$ reduced on $S^1/Z_2$, whose couplings 
to the brane fields are ruled by a Bianchi identity of the schematic form
\be
d C = - {\rm tr} (F \wedge F) \delta(y - y_0) \,.
\label{BianchiC}
\ee
Here and in the following, we shall implicitly understand the splitting of the charged fields over  
the two brane sectors located at different positions $y_0$, but for notational simplicity we shall 
not display this explicitly in the formulae.

\subsection{General structure}

The content of light bosonic fields of the 5D supergravity theory obtained by compactifying 11D supergravity 
on a Calabi-Yau manifold $X$ consists of $1$ symmetric tensor from $G_{MN}$, $h^{1,1}$ scalars from the 
$(1,1)$ components of $G_{i \bar \jmath}$, $h^{1,2}$ complex scalars from the $(1,2)$ and $(2,1)$ 
components of $G_{ij}$ and $G_{\bar \imath \bar \jmath}$, $1$ scalar from the dualization of $C_{MNP}$, 
$1$ complex scalar from the $(3,0)$ and $(0,3)$ components of $C_{ijk}$ and $C_{\bar \imath \bar \jmath \bar k}$, 
$h^{1,1}$ vectors from the $(1,1)$ components of $C_{M i \bar \jmath}$ and $h^{1,2}$ complex scalars 
from the $(1,2)$ and $(2,1)$ components of $C_{i \bar \jmath \bar k}$ and $C_{\bar \imath j k}$. In total this yields 
$1$ symmetric tensor, $h^{1,1} + 4\, h^{1,2} + 3$ real scalar fields and $h^{1,1}$ vector fields, which corresponds 
to the bosonic content of $1$ gravitational multiplet ${\cal G}$ and $1$ universal hypermultiplet ${\cal H}$ 
plus $h^{1,1}-1$ vector multiplets ${\cal V}^a$ associated to the harmonic $(1,1)$ forms arising in addition
to the K\"ahler form and $h^{1,2}$ hyper multiplets ${\cal H}^Z$ associated to the harmonic $(1,2)$ forms 
\cite{CCDF,AFT,LOSW}.

When this 5D theory is further compactified on $S^1/Z_2$ and reinterpreted from a 4D viewpoint, one finds 
$N=2$ supersymmetry projected to $N=1$ supersymmetry. To understand the spectrum of neutral fields, one can 
then think in terms of $N=2$ multiplets and figure out their content in terms of $N=1$ multiplets with definite 
$Z_2$ parities. Listing the even and odd multiplets separated by a semicolon, one finds that 
${\cal G} = (G,T^0;\Gamma)$ where $G$ is the gravitational multiplet, $T^0$ a chiral multiplet and $\Gamma$ is a 
spin-$3/2$ multiplet, ${\cal H} = (S;S^c)$ where $S$ and $S^c$ are chiral multiplets,
${\cal V}^a = (T^a;V^a)$ where $T^a$ are chiral multiplets and $V^a$ vector multiplets, and finally 
${\cal H}^Z= (U^Z;U^{cZ})$ where $U^Z$ and $U^{cZ}$ are chiral multiplets. The spectrum of light 
neutral multiplets thus consists of the graviton $G$, the dilaton $S$, the overall volume modulus $T^0$, 
$h^{1,1}-1$ relative K\"ahler moduli $T^a$ and $h^{1,2}$ complex structure moduli $U^Z$. The 
spectrum of charged fields is instead determined as in the weakly coupled heterotic string, except 
that the fields coming from the two $E_8$ factors are now localized at the two different branes at the 
ends of the $S^1/Z_2$ segment. Altogether they fill a number of $N=1$ chiral multiplets $\Phi^{P}$, $\Psi^{K}$ 
and vector multiplets $V^{x}$, in the representations ${\bf R}$, ${\bf \bar R}$ and ${\bf Adj}$ of the gauge group.

\subsection{Effective K\"ahler potential}

The 4D effective effective K\"ahler potential can be determined by performing the reduction of the 11D 
theory on the Calabi-Yau manifold $X$, and then further reducing the resulting 5D theory on $S^1/Z_2$. 
In this case, it is possible to do the last step by using superfields to directly compute the K\"ahler 
potential, rather than working with the components and looking at the bosonic kinetic terms. 
To perform this computation, we shall do the same approximations as in section 2. We shall first 
neglect the effects of higher-derivative corrections to the 11D effective theory and deformations of the 
basic background, and simply consider the reduction of the two-derivative 11D effective 
theory on $X \times S^1/Z_2$. We shall then also discard the effects of massive Kaluza-Klein modes 
on $X$, although we will retain the effects of massive Kaluza-Klein modes on $S^1/Z_2$, which turn out 
to be crucial to understand the contact terms. Correspondingly, we will also make the same assumptions 
as in section 2, namely that the $(1,1)$ forms $c_{PQ}$ associated to composites of two matter fields are 
harmonic and that the quantities $c^A_{PQ}$ are constant topological invariants. Finally, we shall again 
restrict to the K\"ahler moduli $T^A$ and the charged matter fields $\Phi^P$.

The starting point is the 5D intermediate theory, where we retain not only the $Z_2$-even submultiplets 
$T^0$, $T^a$, $\Phi^{P}$, which contain the light 4D moduli and matter modes, but also the $Z_2$-odd 
submultiplets $V^a$, which contain the heavy Kaluza-Klein modes that have non-trivial linear couplings to the other 
fields and therefore need to be properly integrated out. It is convenient to work with $N=1$ superfields 
$T^0$, $T^a$, $\Phi^{P}$ and $V^a$ depending also on the internal coordinate $y$, and integrate out the 
heavy modes associated to the $V^a$ directly at the superfield level and in a clever way, by solving their 
equations of motion by neglecting space-time derivatives to determine their wave-function profile. In the limit 
where gravity is decoupled, this can be done with usual superfields within rigid supersymmetry along the 
lines of \cite{MSS,AGW,MP}, with $T^0$ playing the role of the radion superfield. Taking into account 
gravitational effects is slightly more complicated, but can actually be done in a very similar way by using 
a superconformal superfield formalism within supergravity, where half of the supersymmetry is manifestly 
realized off-shell. This formalism has been developed in \cite{KO,FKO} and further elaborated in \cite{PST,AS}. 
It has the nice feature of allowing to describe the graviphoton $A_M^0$ on the same footing as the other odd 
gauge fields $A_M^a$, and the volume modulus $T^0$ on the same footing as the other K\"ahler moduli $T^a$, 
through vector multiplets $V^A$ and chiral multiplets $T^A$ with $A=0,a$, at the price of introducing also some 
constraints. The relevant 5D Lagrangian turns out to be
\bea
{\cal L}_{\rm 5D}^{\rm local} \b=\b \int \! d^2 \theta \, 
\bigg[\!-\! \frac 14 \,{\cal N}_{AB} (T^A)\, W^{A\alpha} W^B_\alpha
+ \frac 1{48} \, {\cal N}_{ABC} \, \bar D^2 \big(V^A {\raisebox{11pt}{$$}}^{\leftrightarrow} \hspace{-10pt}  
D^\alpha \partial_y V^B\big) W^C_\alpha \bigg] + {\rm c.c.} \nn \\
\b\;\b \,+ \int \! d^4 \theta \, (- 3)\, {\cal N}^{1/3}(J_y^A) \,. \label{L5Dsugra}
\eea
In this expression, the quantity ${\cal N}$ is a norm function playing the role of real prepotential, which 
is identified with the cubic polynomial defined by the intersection numbers $d_{ABC}$ of the Calabi-Yau 
manifold $X$:
\be
{\cal N}(Z^A) = \frac 16 \, d_{ABC} Z^A Z^B Z^C \,.
\ee
The quantity $W^A_{\alpha}$ denotes the usual super-field-strength associated to $V^A$, namely
\be
W^A_{\alpha} = - \frac 14 \bar D^2 D_\alpha V^A \,.
\ee
Finally, the quantity $J_y^A$ is a current defined in terms of the quantities $c^A_{PQ}$ 
characterizing the vector bundle $V$ over $X$ and given by:
\be
J_y^A = - \partial_y V^A + T^A + \bar T^A - c^A_{PQ} \Phi^P \bar \Phi^Q \, \delta(y - y_0) \,.
\label{JyA}
\ee

In the above expressions, the bosonic modes of $T^A$ come from the decomposition of 
the $2$ forms $J$ and $C_y$ with components $iG_{i \bar \jmath}$ and $C_{y i \bar \jmath}$ 
on the basis of harmonic $(1,1)$ forms $\omega_A$, the bosonic modes of $\Phi^P$ 
come from the decomposition of the Lie-algebra-valued $1$ forms $A$, $\bar A$ with 
components $A_i$, $\bar A_{\bar \imath}$ on the basis of harmonic $1$ forms $u_P$, and 
finally the bosonic modes of $V^A$ come from the decomposition of the $2$ forms 
$C_\mu$ with components $C_{\mu i \bar \jmath}$ on the basis $\omega_A$. The 
correct definition of the chiral multiplets in terms of the above modes turns out to be \cite{PS}
\bea
\a\a T^A = \frac 12 \Big(J^A + i C^A_y + c^A_{PQ}  A^P \bar A^Q \delta(y - y_0) \Big) \label{defT}\,, \\
\a\a \Phi^P = A^P \,. \label{defPhi}
\eea 
We see that these definitions reproduce the ones we have introduced in the component derivation 
of section 2 based on the weakly coupled heterotic string when averaged over the extra dimension. 
Here these definitions ensure that the lowest component of $J_y^A$ simply reduces to the metric 
components, as required in order to reproduce an Einstein gravitational kinetic term coming entirely 
from the bulk and not from the branes, whereas the $\theta \sigma^\mu \bar \theta$ component 
of $J_y^A$ correctly reproduces the modified version of the mixed components of the field strength 
implied by the reduction of the Bianchi identity (\ref{BianchiC}):
\bea
\a\a J_y^A| = J^A \,, \label{Jcomp1}\\
\a\a J_y^A|_{\theta \sigma^\mu \bar \theta} = \partial_\mu A_y^A - \partial_y A_\mu^A 
+ i c^A_{PQ} \Phi^P {\raisebox{11pt}{$$}}^{\leftrightarrow} \hspace{-10pt} \partial_\mu \bar \Phi^Q \delta(y- y_0) \,. \label{Jcomp2}
\eea
This provides a nice superfield interpretation on the need for the shift in the definition of the moduli chiral multiplets.

Integrating out the heavy modes of the vector multiplets $V^A$ effectively amounts to replacing the currents 
$J_y^A$ with their zero modes in the term of the action that does not involve the vector fields. This is easy to show 
in the rigid limit, where only the $V^a$ matter \cite{ASmild}, but actually holds true also in the supergravity regime 
where all the $V^A$ appear but suffer from non-trivial constraints \cite{AS}. One finds the following expression, 
written within the usual superconformal superfield formalism,
\be
{\cal L}_{\rm 4D}^{\rm local} = \int \! d^4 \theta \, (- 3)\, {\cal N}^{1/3}(J^A) \,,
\label{4Dapp}
\ee
where now
\bea
\a\a J^A = T^A + \bar T^A - c^A_{PQ} \Phi^P \bar \Phi^Q \,.
\eea
The effective K\"ahler potential can finally be deduced by matching the integrand of this 
expression with $-3\, e^{-K/3}$. This gives $K = - \log {\cal N}(J^A) = - \log V$, which is the same
result as we obtained directly from the heterotic string:
\be
K = - \log \Big[\frac 16 d_{ABC} J^A J^B J^c \Big] \,, \;\;\text{with}\;\; 
J^A = T^A + \bar T^A - c^A_{PQ} \Phi^P \bar \Phi^Q \,.
\ee
A component version of this five-dimensional derivation is also possible, and was presented in \cite{LOW} for 
the particular case where $h^{1,1} = 1$ with standard embedding.

The effective K\"ahler potential for the untwisted sector of orbifold compactifications can be similarly derived from 
an $M$-theory perspective. The only changes are the same as those already emphasized in section 3, namely that
the intersection numbers $d_{ABC}$ and the quantities $c^A_{PQ}$ acquire a simple group-theoretical 
interpretation. Moreover, in this case the forms $c_{PQ}$ are automatically harmonic and the quantities 
$c^A_{PQ}$ are always constant, as already discussed in section 3. Further details on a component version 
of this five-dimensional derivation can be found in \cite{DG,LLN,NOY,L}.

\subsection{Range of validity}

We have seen in the previous subsection that the results derived in section 2 for the low-energy effective K\"ahler 
potential admit a simple 5D interpretation, in which the non-trivial contact terms spoiling the sequestered structure 
arise from the exchange of heavy 4D Kaluza-Klein modes of the light 5D vector multiplets coming from the 
harmonic components of the $M$-theory $3$-form $C$ on $X$. This interpretation was however derived under the 
same restrictive assumptions as in section 2, namely that the forms $c_{PQ}$ are harmonic and that the quantities 
$c^A_{PQ}$ are constants. It is then natural to wonder once again what would be the situation if these assumptions
were to be relaxed.

The relevance of the assumptions about $c_{PQ}$ and $c^A_{PQ}$ within the $M$-theory perspective must
obviously be very similar to that already discussed within the heterotic perspective. But it turns out to offer 
a slightly sharper perspective. The harmonicity of $c_{PQ}$ is as before needed to ensure the trivial
decoupling of heavy neutral modes from pairs of light charged modes. More specifically, we see here that 
when $c_{PQ}$ is not harmonic a direct danger comes from the heavy 5D vector multiplets that arise from the 
non-harmonic components of the $3$ form $C$ on $X$. Indeed, such heavy modes can be brutally truncated 
away only when they are not sourced by light fields, and from the reduction of the solution of the Bianchi 
identity (\ref{BianchiC}) we see that this is the case only when the non-harmonic parts of $C$ describing the heavy 
5D vector modes have no overlap with the forms $c_{PQ}$ describing the composite of two light matter modes, that 
is when $c_{PQ}$ is harmonic. In the opposite case, one would have to properly integrate out these heavy 5D vector 
modes too, and this would give extra contributions to the contact terms in the 4D effective K\"ahler potential. These
additional effects must correspond to the additional terms that would arise in the left-hand side of eq.~(\ref{contr3}) within 
the heterotic perspective. The constancy of $c^A_{PQ}$ is again needed to ensure a simple determination of 
the right definition of the chiral multiplets containing the moduli. More specifically, we see here that for 
moduli-dependent $c^A_{PQ}$ it is not clear how one should modify the definitions (\ref{defT}) and (\ref{defPhi}) 
to arrange that (\ref{Jcomp1}) and (\ref{Jcomp2}) hold true.
  
\section{General structure of the scalar manifold}
\setcounter{equation}{0}

We have seen that for compactifications on both smooth Calabi-Yau manifolds and singular orbifolds the K\"ahler 
potential for the K\"ahler moduli and matter fields takes the same general form, at least under the already 
explained assumptions. We will now study in some more detail the general geometric features of this scalar manifold, 
which will be relevant for the structure of the soft scalar masses induced in the presence of a non-trivial superpotential. 
We will introduce for this purpose a new parametrization of the scalar manifold, which will turn out to be very convenient 
at some special reference point.

\subsection{Canonical parametrization}

The general class of scalar manifolds we want to study is defined by the following K\"ahler potential, which only depends 
on the two symmetric and Hermitian but otherwise arbitrary constants $d_{ABC}$ and $c^A_{PQ}$:
\be
K = - \log \Big[\frac 16 d_{ABC} J^A J^B J^C \Big] \,,\;\; \text{with} \;\; J^A = T^A + \bar T^A - c^A_{PQ} \Phi^P \bar \Phi^Q\,.
\label{Kuniversal}
\ee
The fields $T^A$ and $\Phi^P$ define a specific parametrization of the scalar manifold defined by this K\"ahler potential,
which naturally emerges from string theory. We are however free to make holomorphic change of coordinates as well 
as K\"ahler transformations to define other equivalent parametrizations. It turns out that this freedom can be used to define 
a particularly convenient kind of parametrization. We shall call this the canonical parametrization, because it is a natural 
generalization including the $N=1$ matter sector of the one that was introduced in \cite{GST,CKVDFdWG} for the very special 
manifolds describing the $N=2$ moduli sector.

The main idea is to think of some reference point of particular interest on the scalar manifold, and then to perform a 
field redefinition that allows to simplify things as much as possible around that point. This reference point can for instance 
be thought of as the one defined by the VEVs $\langle T^A \rangle$ and $\langle \Phi^P \rangle$ that the scalar fields 
would eventually acquire in the presence of a non-trivial superpotential. Since our primary goal is to study situations 
where the moduli have sizable VEVs whereas the matter fields have a small VEVs, we shall start by considering the 
situation where
\be 
\langle T^A \rangle \neq 0 \,,\;\; \langle \Phi^P \rangle = 0 \,.
\ee
We may now reparametrize the fields in such a way to simplify the metric and the curvature tensor at such a point. 
To this aim, we shall consider the following linear field redefinitions:
\be
\hat T^A = U^A_{\;\;B} T^B \,,\;\; \hat \Phi^P = V^P_{\;\;Q} \Phi^Q \,.
\ee
In addition, we may also perform a K\"ahler transformation on $K$. In particular, we may perform a trivial constant shift
of the type
\be
\hat K = K - \log |\alpha|^2 \,.
\ee 
For our purposes, it will be enough to take $U^A_{\;\;B}$ to be a real matrix, $V^P_{\;\;Q}$ to be a complex matrix, 
and $\alpha$ to be a real number. Under such transformations, the new K\"ahler potential in terms of the new fields has the 
same form as the original K\"ahler potential in terms of the original fields, but with new numerical coefficients 
given by:
\be
\hat d_{ABC} = \alpha^2 \, U_{\;\;\;\;\;A}^{\text{-1}D} U_{\;\;\;\;\;B}^{\text{-1}E} U_{\;\;\;\;\;C}^{\text{-1}F} d_{DEF} \,,\;\;
\hat c^A_{PQ} = U^A_{\;\;B} V_{\;\;\;\;\;\;P}^{\text{-1}R} \bar V_{\;\;\;\;\;Q}^{\text{-1}S} c^B_{RS} \,.
\ee
At this point, we may choose $U^A_{\;\;B}$ and $V^P_{\;\;Q}$ in such a way that the 
VEVs of the fields are aligned along just one direction, the VEV of the metric becomes 
diagonal, and the overall scale of one of these two quantities (but not both) is set to 
some reference value. We may furthermore choose $\alpha$ to set the overall scale 
of the intersection numbers to a convenient value. More specifically, we shall 
require that in the new basis the reference point should be at
\be
\langle \hat T^A \rangle = \frac {\sqrt{3}}2 \, \delta^A_0 \,,\;\; \langle \hat \Phi^P \rangle = 0 \,,
\ee
the metric at that point should take the form
\be
\langle \hat g_{AB} \rangle = \delta_{AB} \,,\;\;
\langle \hat g_{PQ} \rangle = \delta_{PQ}\,,\;\;
\langle \hat g_{AQ} \rangle = 0 \,,
\ee
and finally the K\"ahler frame should be such that at that point
\be
\langle \hat K \rangle = 0 \,.
\ee
It is easy to get convinced by a counting of parameters that it is indeed always possible to 
impose this kind of conditions. Moreover, by comparing the transformed expressions for the 
VEVs of the fields, the metric and the K\"ahler potential with the values required in the previous 
equations, we deduce that the new values of the numerical coefficients $\hat d_{ABC}$ and 
$\hat c^A_{PQ}$ must satisfy the following properties:
\bea
\a\a \hat d_{000} = \frac 2{\sqrt{3}} \,,\;\; \hat d_{00a} = 0 \,,\;\; 
\hat d_{0 ab} = - \frac 1{\sqrt{3}}\, \delta_{ab} \,,\;\; \hat d_{abc} = \text{generic} \,, \label{dcan} \\
\a\a \hat c^0_{PQ} = \frac 1{\sqrt{3}} \, \delta_{PQ} \,,\;\; \hat c^a_{PQ} = \text{generic} \label{lambdacan} \,.
\eea
The new form of the K\"ahler potential after the change of basis is then
\be
\hat K = - \log \bigg[\frac 16 \Big(\frac 2{\sqrt{3}} \hat J^0 \hat J^0 \hat J^0 - \sqrt{3}\, \hat J^0 \hat J^a \hat J^a 
+ \hat d_{abc} \hat J^a \hat J^b \hat J^c \Big) \bigg]\,,
\ee
where now
\bea
\a\a \hat J^0 = \hat T^0 + \hat {\bar T}^0 - \frac 1{\sqrt{3}} \delta_{PQ} \hat \Phi^P \hat {\bar \Phi}^Q \,, \\
\a\a \hat J^a = \hat T^a + \hat {\bar T}^a - \hat c^a_{PQ} \hat \Phi^P \hat {\bar \Phi}^Q \,.
\eea

The above canonical parametrization has a nice interpretation from the point of view of the properties of the 
Calabi-Yau manifold $X$ and the holomorphic vector bundle $V$ over it, on which the model is based. It 
essentially corresponds to a particular choice of bases for the harmonic forms $\hat \omega_A$ and $\hat u_P$ 
at the reference point defined by the VEVs. More specifically, the sets of harmonic forms $\hat \omega_A$ and 
$\hat u_P$ can be chosen to be orthonormal with respect to the natural positive definite metrics defined by 
$\hat g_{AB} = V^{-1} \! \int_X \hat \omega_A \wedge * \hat \omega_B$ and $\hat g_{PQ} = V^{-1} \! \int_X \hat c_{PQ} \wedge *J$, 
and one can moreover orient them in such a way that $\hat \omega_0$ is aligned with the K\"ahler form $J$. In this 
way the multiplets $\hat T^0$ and $\hat T^a$ describe respectively the overall volume and the relative K\"ahler
moduli, and the fields $\hat \Phi^P$ are canonically defined. In this new basis, the VEV of the metric is the identity
matrix, with $\hat g_{AB} = \delta_{AB}$ and $\hat g_{PQ} = \delta_{PQ}$, and as shown in appendix A the intersection 
numbers $\hat d_{ABC}$ and the quantities $\hat c^A_{PQ}$ do indeed take the structure of (\ref{dcan}) and 
(\ref{lambdacan}), after effectively setting the volume $V$ to unity by a rescaling. It is worth remarking that 
if the traceful part of $\hat c_{PQ}$ were parallel to $J$ and thus proportional to $\hat \omega^0$, whereas the 
remaining traceless part of $\hat c_{PQ}$ were orthogonal to $J$ and thus a linear combination of 
the $\hat \omega^a$'s, all the matrices $\hat c^a_{PQ}$ would be traceless. This turns out to be the case 
for orbifolds, and it is not unconceivable that it might actually also hold true for most if not all of the Calabi-Yau's 
subject to the stringent restriction that the $(1,1)$ forms $c_{PQ}$ are harmonic. We were not able to verify this,
but we find it rather suggestive that the trace part of $\hat c_{PQ}$ indeed has positive-definite components, like
$J$.

Notice that the new coordinates that have been introduced do not exactly coincide with normal coordinates at the 
reference point. Indeed, some of the components of the Christoffel connection have non-trivial values:
\bea
\a\a \langle \Gamma_{00\bar 0} \rangle = - \frac 2{\sqrt{3}} \,,\;\;
\langle \Gamma_{0 a \bar b} \rangle = - \frac 2{\sqrt{3}} \delta_{ab} \,,\;\;
\langle \Gamma_{ab \bar 0} \rangle = - \frac 2{\sqrt{3}} \delta_{ab} \,,\;\;
\langle \Gamma_{ab \bar c} \rangle = - \hat d_{abc} \,, \\
\a\a \langle \Gamma_{AP\bar Q} \rangle = - \hat c^A_{PQ} \,.
\eea
Nevertheless, they turn out to lead to rather simple expressions for the Riemann curvature tensor at the reference 
point.

\subsection{Curvature for Calabi-Yau models}

In the general case of compactifications on a smooth Calabi-Yau manifold, the scalar manifold ${\cal M}$ on 
which the low-energy effective theory is based is a generic K\"ahler manifold.
The curvature of such a manifold depends on the point. Let us then consider the special reference point 
introduced above, assuming that it is dynamically selected by the superpotential, and let us switch to the canonical
parametrization. After a simple computation, one finds the following results for the VEV of the Riemann tensor:
\bea
\a\a \langle R_{A \bar B C \bar D}\rangle = \delta_{AB} \delta_{CD} + \delta_{AD} \delta_{BC} - \hat d_{ACE} \hat d_{BDE} \,, \\[1.7mm]
\a\a \langle R_{P \bar Q R \bar S} \rangle = 
\frac 13 \big(\delta_{P Q} \delta_{R S} + \delta_{PS} \delta_{RQ} \big) + \hat c^a_{P Q} \hat c^a_{R S} + \hat c^a_{P S} \hat c^a_{R Q} \,, \\
\a\a \langle R_{\hspace{-1pt} P \bar Q 0 \bar 0} \rangle = \frac 13 \delta_{\hspace{-1pt} P Q} \,,\;\,
\langle R_{\hspace{-1pt} P \bar Q a \bar b} \rangle = \frac 23 \delta_{\hspace{-1pt} P Q} \delta_{ab} 
\!+\! (\hat d_{abc} \hat c^c \!-\! \hat c^a \hat c^b)_{\hspace{-1pt}P Q} \,,\;\,
\langle R_{\hspace{-1pt} P \bar Q 0 \bar b} \rangle = \frac 1{\!\sqrt{3}} \hat c^b_{\hspace{-1pt} P Q} \,.
\eea
These expressions are valid only around the point under consideration. In particular, they get deformed if 
one switches on a non-vanishing VEV for the matter fields.

\subsection{Curvature for orbifold models}

In the special case of orbifold compactifications, the scalar manifold ${\cal M}$ on which the low-energy effective 
theory is based is a symmetric K\"ahler manifold. The curvature of such a manifold does not depend on the point. 
Let us nevertheless consider the special reference point introduced above and switch as before to the canonical 
parametrization. It is straightforward to verify that the new parametrization described in section \ref{GS} actually 
coincides with the canonical one. To do so, one simply needs to recall that $c^0$ is equal to $1\!\! 1/\sqrt{3}$, 
whereas the $c^a$ are a subset of the transposed of the Gell-Mann matrices $\lambda^a$. One then verifies that the expressions 
(\ref{dorb}) and (\ref{lambdaorb}) do indeed take the canonical forms defined by (\ref{dcan}) and (\ref{lambdacan}),
with:
\bea
\hat d_{abc} = 2\, {\rm tr}(\lambda^{(a} \lambda^b \lambda^{c)}) \,,\;\; \hat c^a_{ij} = \lambda^a_{ji} \,.
\eea
We see that in this case $\hat d_{abc}$ is the symmetric invariant symbol of the group $H$, whereas the 
$\hat c^a_{ij}$ are the transposed of the generators of $H$ in the representation ${\bf h}$ descending from 
the ${\bf 3}$ of $SU(3)$ in terms of $3 \times 3$ matrices. In this case the transposed of the matrices 
$\hat c^a_{ij}$ possess the non-trivial property of being traceless and generating the Lie algebra of $H$, 
whose structure constants can be written as
\be
f_{abc} = - 2i\, {\rm tr}(\lambda^{[a} \lambda^b \lambda^{c]}) \,.
\ee
Moreover, for all the three kinds of models one finds:
\be
[\lambda^a,\lambda^b] = i f_{abc} \lambda^c \,,\;\;
\{\lambda^a,\lambda^b\} = d_{abc} \lambda^c + \frac 23 \delta_{ab} 1\!\!1 \,.
\ee
Using these properties of the matrices $\lambda^a$, the components of the Riemann tensor are then 
seen to simplify and can entirely be rewritten in terms of these matrices:
\bea
\a\a \langle R_{A \bar B C \bar D} \rangle = {\rm tr} (\hat c^A \hat c^B \hat c^C \hat c^D) + {\rm tr} (\hat c^A \hat c^D \hat c^C \hat c^B)\,, \\[0.5mm]
\a\a \langle R_{P \bar Q R \bar S} \rangle = \hat c^A_{PQ} \hat c^A_{RS} + \hat c^A_{PS} \hat c^A_{RQ} \,, \\
\a\a \langle R_{P \bar Q C \bar D} \rangle = (\hat c^D \hat c^C)_{PQ} \,.
\eea
These expressions are actually valid at any point of the scalar manifold, as already said. Their simple form 
reflects the fact that the curvature of symmetric manifolds is completely determined by the structure constants
of their isometry group. This is explained in some detail in appendix B, where we also summarize some basic 
results about the geometry of such symmetric coset manifolds.

\section{Soft scalar masses and sequestering}
\setcounter{equation}{0}

Let us now come to the crucial question of what are the properties of soft scalar masses in the effective theories 
for heterotic string models compactified on a generic Calabi-Yau manifold with a generic stable holomorphic vector 
bundle over it, in the presence of some source of supersymmetry breaking. We shall restrict our analysis to the 
K\"ahler moduli and matter fields, for which we know the form of the K\"ahler potential, and to the neighborhood 
of the reference point introduced last section, by assuming that the superpotential that induces supersymmetry 
breaking is such that the scalar VEVs of the moduli and matter scalar fields are respectively generic and vanishing.
We will first work out the general structure of the soft scalar masses and then study the possibility of ensuring the 
vanishing of these masses with the help of some kind of global symmetry. 

\subsection{Structure of scalar masses}

Our starting point is the effective K\"ahler potential (\ref{Kuniversal}), which is characterized by the two constants
$d_{ABC}$ and $c^A_{PQ}$. Since we want to study soft terms at the particular reference point introduced in last 
section, it will be convenient to switch to the canonical parametrization that we defined there. From now on, we 
shall for simplicity drop all the hats on the redefined parameters and fields, and also the brackets denoting VEVs
at the reference point. It will moreover be convenient to further redefine $T = T^0/\sqrt{3}$ and correspondingly 
$J = J^0/\sqrt{3}$, and to explicitly split the matter fields $\Phi^P$ into two sets $Q^\alpha$ and $X^i$ 
respectively coming from the two $E_8$ factors, in such a way to match the notation that was adopted in \cite{ASmild} 
for orbifold models. The visible sector is then identified with the fields $Q^\alpha$ and the hidden sector generically
contains all the remaining fields $X^i, T, T^a$, and the K\"ahler potential becomes
\be
K = - \log \Big(J^3 - \frac 12 J J^a J^a + \frac 16 d_{abc} J^a J^b J^c \Big) \,,
\label{Kconv}
\ee
where
\bea
\a\a J = T + \bar T - \frac 13 Q^\alpha \bar Q^\alpha - \frac 13 X^i \bar X^i \,, \\
\a\a J^a = T^a + \bar T^a - c^a_{\alpha \beta} Q^\alpha \bar Q^\beta - c^a_{ij} X^i \bar X^j  \,.
\eea
Let us now study this expression around the point under consideration. In the new coordinates, this
corresponds to:
\be
T = \frac 12 \,,\;\; T^a = 0 \,,\;\; Q^\alpha = 0 \,,\;\; X^i = 0 \,.
\ee
The metric takes a simple diagonal result, with non-vanishing entries given by
\be
g_{T \bar T} = 3 \,,\;\; g_{a \bar b} = \delta_{ab} \,,\;\;
g_{\alpha \bar \beta} = \delta_{\alpha \beta} \,,\;\; 
g_{i \bar \jmath} = \delta_{ij} \,.
\label{metricref}
\ee
For the Christoffel connection, the non-vanishing components are given by
\bea
\a\a \Gamma_{T T \bar T} = - 6 \,,\;\;
\Gamma_{T a \bar b} = - 2\, \delta_{ab} \,,\;\;
\Gamma_{ab \bar T} = - 2\, \delta_{ab} \,,\;\;
\Gamma_{ab \bar c} = - d_{abc} \,, \\[1mm]
\a\a \Gamma_{T P\bar Q} = - \delta_{PQ} \,,\;\;
\Gamma_{aP\bar Q} = - c^a_{PQ} \,.
\eea
The components of the Riemann tensor that are relevant for soft scalar terms, with a pair of indices 
along the visible sector fields and the other pair along the hidden sector fields, are then found to be
\bea
\a\a R_{\alpha \bar \beta i \bar \jmath} =
\frac 13 \delta_{\alpha \beta} \delta_{ij} + c^a_{\alpha \beta} c^a_{ij} \,, \label{curvref1} \\
\a\a R_{\alpha \bar \beta T \bar T} = \delta_{\alpha \beta} \,,\;\;
R_{\alpha \bar \beta a \bar b} = \frac 23 \delta_{\alpha \beta}\delta_{ab} 
+ (d_{abc} c^c \!- c^a c^b)_{\alpha \beta} \,,\;\;
R_{\alpha \bar \beta T \bar b} = c^b_{\alpha \beta} \,. \label{curvref2}
\eea

We are now in position to compute the soft scalar masses induced for the visible-sector fields $Q^\alpha$ 
when the hidden-sector fields $\Phi^\Theta = X^i, T, T^a$ get non-vanishing auxiliary fields, at the reference 
point under consideration. This can be done by using the following standard geometrical expression 
\be
m^2_{\alpha \bar \beta} = - \Big(R_{\alpha \bar \beta \Theta \bar \Gamma}
- \frac 13 g_{\alpha \bar \beta} g_{\Theta \bar \Gamma} \Big) F^\Theta \bar F^{\bar \Gamma} \,.
\ee
Using the results (\ref{metricref}) and (\ref{curvref1})-(\ref{curvref2}) for the metric and the Riemann tensor 
at the point under consideration, this gives:
\bea
m^2_{\alpha \bar \beta} \b=\b - c^a_{\alpha \beta} c^a_{i j} F^i \bar F^{\bar \jmath} 
- \Big(\frac 13 \delta_{\alpha \beta}\delta_{ab} + (d_{abc} c^c \!- c^a c^b)_{\alpha \beta}\Big) F^a \bar F^{\bar b} \nn \\
\b\;\b -\, c^a_{\alpha \beta} F^a \bar F^T + {\rm c.c.} \,.
\label{msoft}
\eea

The structure of the soft scalar masses (\ref{msoft}) can also be understood in terms of ordinary superfields. 
To do this, one considers the kinetic function $\Omega = -3\, e^{-K/3}$, which is the gravitational analogue of the 
rigid K\"ahler potential. At the considered reference point, it is sufficient to expand it at cubic order in 
$J^a \ll J$. In this way one finds:
\bea
\Omega \b\simeq\b - 3\, J + \frac 12 \frac {J^a J^a}{J} - \frac 16 d_{abc} \frac {J^a J^b J^c}{J^2} \,.
\label{Omega}
\eea
More precisely, the relevant terms are selected by decomposing the fields in scalar VEVs plus fluctuations, 
so that $J = 1 + \tilde J$ and $J^a = \tilde J^a$, and retaining up to cubic terms in an expansion in powers 
of the fluctuations. This yields $\Omega = - 3 + \tilde \Omega$ with:
\bea
\tilde \Omega \b\simeq\b - 3\, \tilde J + \frac 12 \tilde J^a \tilde J^a - \frac 12 \tilde J \, \tilde J^a \tilde J^a 
- \frac 16 d_{abc} \tilde J^a \tilde J^b \tilde J^c \,.
\label{Omegatilde}
\eea
The soft scalar masses can the be computed by looking at the quadratic part of the contribution to
the scalar potential from $\tilde \Omega$: ${\cal L}_{m^2} = - \tilde \Omega|_{D,q^2}$. The various terms 
in (\ref{msoft}) then emerge as follows from $\tilde \Omega|_D$, after splitting the currents into visible-sector 
and hidden-sector parts. The term $- c^a_{\alpha \beta} c^a_{i j} F^i \bar F^{\bar \jmath}$ 
comes from $\tilde J^a_{\rm v}| \tilde J^a_{\rm h}|_D$, 
the term $-1/3 \, \delta_{\alpha \beta}\delta_{ab} F^a \bar F^{\bar b}$ comes from
$- \tilde J_{\rm v}| \tilde J^a_{\rm h}|_F \tilde J^a_{\rm h} |_{\bar F}$, 
the term $- c^a_{\alpha \beta} F^a \bar F^T + {\rm c.c.}$ comes from 
$- \tilde J_{\rm h}|_{\bar F} \tilde J_{\rm v}^a| \tilde J^a_{\rm h} |_{F} + {\rm c.c.}$, 
the term $(c^a c^b)_{\alpha \beta} F^a \bar F^{\bar b}$ 
comes from the combination of $-3\,\tilde J_{\rm v}|_D$ and 
$\tilde J^a_{\rm v}|_F \tilde J^a_{\rm h}|_{\bar F} + {\rm c.c.}$, and 
finally the term $-d_{abc} c^a_{\alpha \beta} F^b \bar F^{\bar c}$ comes from 
$- d_{abc} \tilde J_{\rm v}^a| \tilde J_{\rm h}^b|_F \tilde J^c_{\rm h} |_{\bar F}$.
 
\subsection{Sequestering by global symmetries}

From the form of the expression (\ref{msoft}), we can deduce the following observations. In the particular 
case where $h^{1,1} = 1$, the soft scalar masses vanish identically, even in the presence of generic non-vanishing 
values for $F^T$ and $F^i$. This is the well known situation arising in sequestered models. In the general case 
where $h^{1,1} > 1$, one the contrary, the soft scalar masses receive non-trivial contributions in the presence of 
generic non-vanishing values of $F^T$, $F^i$ and $F^a$. However, these contributions involve very special 
combinations of these auxiliary fields, controlled by the quantities $d_{abc}$ and the matrices $c^a_{\alpha \beta}$ 
and $c^a_{ij}$. One may then wonder whether it is possible to ensure that these combinations of auxiliary 
fields vanish, so that the soft scalar masses would again vanish, by assuming that some approximate global 
symmetry of the K\"ahler potential $K$ is extended to constrain also the superpotential $W$ and therefore the 
Goldstino direction. It would also be interesting to study what constraints are put on the Goldstino 
direction by the requirement that there should exist a metastable supersymmetry breaking vacuum, 
generalizing the results derived in \cite{CGGLPS} for K\"ahler moduli to include also matter fields,
but we shall not attempt to do this here.

From the results derived in the previous subsection, and taking into account that the scalar VEVs of the fields 
$T^a$ and $X^i$ are assumed to be negligible, we see that a simple and general possibility to get vanishing 
soft scalar masses is to require that:
\bea
\a\a c^a_{ij} F^i \bar F^{\bar \jmath} = 0 \;\Leftrightarrow\; J^a_{\rm h}|_D = 0 \,, \label{relD}\\
\a\a F^a = 0 \;\Leftrightarrow\; J^a_{\rm h}|_F = 0 \label{relF} \,.
\eea
These two relations clearly have the form of the two $D$ and $F$ type Ward identities that would be implied by 
the conservation of the currents 
\be
J_{\rm h}^a = T^a + \bar T^a - c^a_{ij} X^i \bar X^j \,.
\label{Jh}
\ee
Notice however that one might also view the two relations (\ref{relD}) and $(\ref{relF})$ as emerging from 
the conservation of the following two independent currents, which each lead to only one non-trivial Ward 
identity, respectively the $D$ and $F$ type one:
\bea
\a\a J_{{\rm h}X}^a = - c^a_{ij} X^i \bar X^j \,, \label{JhX} \\
\a\a J_{{\rm h}T}^a = T^a + \bar T^a \,. \label{JhT}
\eea
This follows form the observation that at the considered vacuum reference point one finds
$J_{{\rm h}X}^a|_D = J_{{\rm h}}^a|_D$, $J_{{\rm h}X}^a|_F = 0$, 
$J_{{\rm h}T}^a|_D = 0$ and $J_{{\rm h}T}^a|_F = J_{{\rm h}}^a|_F$.   

To understand which global symmetry would lead to this conserved current, let us now recall that the 
general form of the conserved N\"other current superfield $J^a$ for a globally supersymmetric 
non-linear sigma model with a global symmetry $\delta \Phi^I = k_a^I \delta \epsilon^a$ is given, in  
the general case where the K\"ahler potential is allowed to undergo a K\"ahler transformation parametrized
by some holomorphic functions $f_a$, by the following expression:
\be
J^a = {\rm Im} (k_a^I K_I  - f_a) \,.
\label{JN}
\ee 
The $D$ and $F$ type Ward identities following from the conservation of this current 
take the following form:
\bea
\a\a J^a|_D = 0 \;\Leftrightarrow\; \nabla_I k_{a \bar J} F^I \bar F^{\bar J} = 0 \,, \label{WardD} \\
\a\a J^a|_F = 0 \;\Leftrightarrow\; \bar k_{aI} F^I = 0 \, \label{WardF}
\eea
Somewhat surprisingly, gravitational effects complicate the situation \cite{ASmild}. Although 
it is not totally trivial to generalize the superfield expression (\ref{JN}), it is rather straightforward 
to show that the two component Ward identities (\ref{WardD}) and (\ref{WardF}) are deformed 
to $\nabla_I k_{a \bar J} F^I \bar F^{\bar J} = - 2i D_a m_{3/2}^2$ and $k_{aI} F^I = - i D_a m_{3/2}$, 
where $D_a = {\rm Im}(k_a^I K_I - f_a)$. This is due to the fact that the auxiliary fields $F^I$ receive 
a gravitational contribution involving derivatives of $K$, in addition to the usual contribution involving 
derivatives of $W$. Notice however that at the particular reference point that we have considered,
the only non-vanishing component of $K_I$ is along the $T$ direction, so that $K_\alpha = 0$,
$K_i = 0$ and $K_a = 0$. Under the mild restriction that the considered symmetry should not act 
on $T$ and should not involve a K\"ahler transformation, meaning that $k_a^T = 0$ and $f_a=0$,
one would then get $D_a = 0$. Under this assumption, one can then use the rigid version of 
the Ward identities.

To get an idea of the situation, we may now start by naively applying the expression (\ref{JN}) 
with a K\"ahler potential $K$ given by the leading quadratic part of $\Omega$, namely
\be
K \simeq  \frac 12 (T^a + \bar T^a)(T^a + \bar T^a) + X^i \bar X^i \,.
\label{Kleading}
\ee
To match (\ref{JN}) with the two partial currents (\ref{JhX}) and (\ref{JhT}), we would then 
respectively need to take $k_a^i \simeq - i c^a_{ji} X^j$ for the matter fields $X^i$ 
and $k_a^b \simeq i \delta_a^b$ for the moduli fields $T^a$. These Killing vectors define 
two sets of transformations that indeed leave the leading K\"ahler potential (\ref{Kleading}) 
independently invariant:
\bea
\a\a \delta_a X^i \simeq - i c^a_{ji} X^j \,, \label{dX} \\
\a\a \delta_a T^b \simeq i \delta_a^b \,. \label{dT}
\eea
The crucial question is now whether the transformations (\ref{dX}) and (\ref{dT}) are eligible to represent an 
approximate global symmetry of $K$ around the vacuum reference point under consideration or not. 
A first condition is that the matrices $c^a$ should form a closed algebra with 
$[c^a,c^b] = - i f_{abc} c^c$. In this way the transformations (\ref{dX}) would form an 
algebra with structure constants $f_{abc}$ associated to a group $H$, while the transformations (\ref{dT}) 
automatically form an Abelian algebra associated to $U(1)^{h^{1,1}-1}$. 
A second condition is that higher order terms in $K$ should have an unimportant effect and that it should 
somehow be meaningful to impose to $W$ a symmetry that leaves a priori invariant only the leading quadratic 
part of $K$. One possibility is that the corrections spoil the symmetries (\ref{JhX}) and (\ref{JhT}) but 
only in a parametrically suppressed way. It is however not clear whether this can robustly happen. A more 
appealing possibility is that (\ref{dX}) and (\ref{dT}) can be extended to exact symmetries of the full scalar 
manifold, thereby guaranteeing the existence of exactly conserved currents which reduce to (\ref{JhX}) 
and (\ref{JhT}) in the vicinity of the point under consideration. We see however from the form (\ref{Kconv}) 
of $K$ that (\ref{dX}) can be generalized to an exact symmetry only by extending it to act linearly also on 
the $T^a$ in the adjoint representation of $H$ and only if $d_{abc}$ corresponds to an invariant of the group 
$H$, while (\ref{dT}) is always an exact symmetry, without the need of any modification and for any values of 
$d_{abc}$. The exact conserved currents differ from (\ref{JhX}) and (\ref{JhT}), on one hand because of the 
extension in the symmetry action and on the other because of the non-linearities in the K\"ahler potential. 
The Ward identities (\ref{WardD}) and (\ref{WardF}) are then correspondingly deformed. However, taken 
together they still ensure that $c^a_{ij}F^i \bar F^{\bar \jmath} = 0$ and $F^a = 0$, which guarantee the 
vanishing of the soft scalar masses.

In addition to the general possibility that we just explored, there might also be other options that arise in 
specific situations. For instance, the three terms of the second piece in (\ref{msoft}) may conspire to give 
a simpler structure, and one might try to exploit this in the search for a different global symmetry that could 
ensure the vanishing of soft masses by constraining the $F^a$'s but without setting them all to zero. In such 
a case one would however have to assume that $F^T$ vanishes to get rid of the last piece in (\ref{msoft}).
Let us then study more specifically what are the options for general Calabi-Yau models and for orbifold models, 
focusing for simplicity on models with a symmetric embedding in the visible and hidden sectors, for which the 
set of matrices $c^a_{\alpha \beta}$ and $c^a_{ij}$ are identical.

\subsection{Calabi-Yau models}

For generic Calabi-Yau models, the intersection numbers $d_{abc}$ and the Hermitian matrices 
$c^a_{\alpha\beta}$ or equivalently $c^a_{ij}$ are a priori generic, with $a=1,\cdots,h^{11}-1$ and 
$\alpha,\beta,i,j=1,\cdots,n_R$. The only thing that we know for sure from the discussion of section 2.3 
is that the matrices $c^a$ and $c^0$ can always be written as transposed linear combinations of 
the $n_R^2$ matrices $\lambda^{A'}$ representing the generators of $U(n_R)$ in the fundamental 
representation. As remarked at the end of section 5, a further property that could conceivably 
arise with some naturalness and generality is that these matrices might be traceless. In that case
they could then be expressed in terms of the $n_R^2-1$ traceless generators of $SU(n_R)$.  
On the other hand, further restrictions leading to yet smaller subgroups $H'$ seem less 
likely, and the minimal case where the matrices $c^a$ themselves generate a group $H$
of dimension $h^{1,1}-1$ appears to be very special.

Consider first the brane-mediated effect corresponding to the first term of (\ref{msoft}). If the matrices 
$c^a$ happen to be transposed linear combinations of the generators $\lambda^{a'}$ of some 
group $H' \subset U(n_R)$, we may ensure the vanishing of this contribution by imposing 
the global symmetry $H'$ that acts as in (\ref{dX}) but with $c^a_{ji}$ replaced by $\lambda^{a'}_{ij}$: 
$\delta_{a'} X^i = - i\, \lambda^{a'}_{ij} X^j$. 
This is still an approximate symmetry of $K$ and leads to the conservation of the larger set of currents 
$J^{a'}_{{\rm h}X} = - \lambda^{a'}_{ji} X^i \bar X^j$, which implies the stronger Ward identity 
$\lambda^{a'}_{ji} F^i \bar F^j = 0$. The maximal choice $H'=U(n_R)$ is available for any generic model, 
but has the drawback that it would actually imply $F^i = 0$, due to the completeness relation 
$\lambda^{a'}_{ij} \lambda^{a'}_{pq} = \delta_{iq} \delta_{pj}$. Other non-maximal choices 
$H' \subset U(n_R)$ are instead available only in particular models, but have the advantage of 
allowing $F^i \neq 0$. 
Notice finally that such an approximate symmetry group $H'$ can in general not be extended 
to an exact symmetry of the full scalar manifold. The only very special case where this is possible
is when the $c^a$ generate by themselves a minimal group $H$ of dimension $h^{1,1}-1$
and the intersection numbers $d_{abc}$ are invariant under this group $H$. 

Consider next  the moduli-mediated effect corresponding to the remaining terms of (\ref{msoft}). 
In general one may ensure that these vanish by imposing the independent Abelian global symmetry 
$U(1)^{h^{1,1}-1}$ acting as in (\ref{dT}): $\delta_a T^b = i\, \delta_a^b$. This symmetry leads to 
the conservation of the currents $J^a_{{\rm h}T} = T^a + \bar T^a$, and the corresponding $F$ 
type Ward identity implies that $F^a = 0$. Moreover it always corresponds to an exact symmetry 
of the full scalar manifold. Notice finally that in this case it is rather unlikely that the second piece 
of (\ref{msoft}) could simplify dramatically enough to allow for other options.

We conclude that for smooth Calabi-Yau compactifications there generically exists the possibility 
of ensuring the vanishing of soft scalar masses at points with negligible VEVs for $X^i$ and $T^a$ by 
imposing the approximate global symmetry $U(n_R) \times U(1)^{h^{1,1}-1}$, where the first factor 
acts linearly on the $X^i$ and the second acts as a shift on the $T^a$. However, this forces both the 
$F^i$ and the $F^a$ to vanish, meaning that there is actually no breaking of supersymmetry at all.
Moreover, it is not a true symmetry of the full scalar manifold. A more interesting situation may be 
obtained in the special cases where the matrices $c^a$ generate some 
non-maximal subgroup $H \subset U(n_R)$. In such a situation, the $F^i$ would be constrained 
but not forced to vanish, although the $F^a$ would still vanish, and supersymmetry can be broken. 
Moreover, this symmetry can be extended to a true symmetry of the full scalar manifold that still 
implies the vanishing of the scalar masses.

\subsection{Orbifold models}

For orbifold models, the intersection numbers $d_{abc}$ and the matrices $c^a_{\alpha\beta}$ or 
equivalently $c^a_{ij}$, with $a = 1,\cdots,h^{1,1}-1$ and $\alpha,\beta,i,j = 1,2,3$, are a respectively 
the symmetric invariant symbol and the transposed tridimensional representation of the generators 
of a group $H \subset SU(3)$. Moreover, one can easily verify that the second term in (\ref{msoft}) simplifies to 
$1/3\, \delta_{\alpha \beta}\delta_{ab} + (d_{abc} c^c \!-\! c^a c^b)_{\alpha \beta}
= (c^b c^a)_{\alpha \beta} - 1/3\, \delta_{ab} \delta_{\alpha \beta}$,
which is traceless. As a result, the mass matrix (\ref{msoft}) is traceless and depends only on $h^{1,1}-1$ 
independent parameters, which can be taken to be $c^a_{ji} m^2_{ij}$.

Consider first the first brane-mediated term in (\ref{msoft}). In this case, this can be ensured to vanish 
by imposing the global symmetry $H$ acting as in (\ref{dX}): $\delta_a X^i = - i \lambda^a_{ij} X^j$.
This leads to the conservation of the currents $J^a_{{\rm h}X} = - \lambda^a_{ji} X^i \bar X^j$, which implies 
the $D$ type Ward identity $\lambda^a_{ji} F^i \bar F^j = 0$. Moreover, this approximate symmetry can 
be extended to an exact symmetry of the full manifold, as explained in appendix B, by assigning a non-trivial 
linear transformation law to the fields $T^a$ in the adjoint representation of $H$. Notice finally that in this case 
one does not have the option of enlarging the symmetry to a bigger group $H' \subset U(n_R)$, because 
the various generations are grouped into triplets transforming in the fundamental representation of the gauge 
group enhancement factor, which happens to coincide with $H$.

Consider next the remaining moduli-mediated terms in (\ref{msoft}). In general, we may again ensure the 
vanishing of these terms by imposing an independent Abelian global symmetry $U(1)^{h^{1,1}-1}$ 
acting as in (\ref{dT}): $\delta_a T^b = i\, \delta_a^b$. This leads to the conservation of the currents 
$J^a_{{\rm h}T} = T^a + \bar T^a$, which implies the $F$ type Ward identity $F^a = 0$. Moreover, this 
symmetry is actually as before an exact symmetry of the full scalar manifold. Notice finally that in this 
case the second piece of (\ref{msoft}) actually simplifies to $(d_{abc} + i f_{abc}) F^b \bar F^c$. One 
may then wonder whether the vanishing of this moduli-mediated contribution could perhaps be achieved 
together with the brane-mediated contribution with a single exact global symmetry $H$, acting on both the 
$X^i$ and the $T^a$ respectively in the fundamental and in the adjoint representations. Comparing with 
the structure (\ref{WardD}) of the Ward identity, we however see that this does not work.

We conclude that for toroidal orbifold compactifications there always exists the possibility of ensuring the 
vanishing of soft scalar masses at points with negligible VEVs for $X^i$ and $T^a$ by imposing the 
approximate global symmetry $H \times U(1)^{h^{1,1}}$, where the first factor acts linearly on the $X^i$ 
and the second factor acts as a shift on the $T^a$. In this situation, the $F^i$ would be constrained 
but not forced to vanish, although the $F^a$ would still vanish, and supersymmetry can be broken. 
Moreover, this symmetry can be extended to a true symmetry of the full scalar manifold that still 
implies the vanishing of the scalar masses. 

\section{Conclusions}
\setcounter{equation}{0}

In this paper, we have attempted a general study of the structure of soft scalar masses in heterotic string models 
obtained by compactification on a Calabi-Yau manifold $X$ with a stable holomorphic vector bundle $V$ over it. 
We investigated in particular the possibility of ensuring that such masses vanish at the classical level, by an effective 
sequestering mechanism based on global symmetries, and are then dominated by approximately universal 
quantum effects, so that the supersymmetric flavor problem could be naturally solved. Our main goal was to generalize
a similar study previously done in \cite{ASmild} for the special case of singular orbifolds, and to assess how much 
of the structure allowing for an interesting implementation of this mechanism survives in the general case of 
smooth Calabi-Yau manifolds. We focused for simplicity on the low-energy effective theory restricted to the K\"ahler 
moduli $T^A$ and the charged matter fields $Q^\alpha$ and $X^i$ coming from the two $E_8$ sectors, with the 
$Q^\alpha$ defining the visible sector and the $X^i$ and $T^A$ the hidden sector. We then studied the terms in 
the effective K\"ahler potential $K$ that mix the visible matter fields $Q^\alpha$ with either the hidden moduli fields 
$T^A$ or the hidden matter fields $X^i$, and the moduli-mediated and brane-mediated contributions to soft scalar 
masses for $Q^\alpha$ that these operators induce when $T^A$ and $X^i$ acquire some non-vanishing auxiliary 
fields due to a superpotential $W$ of unspecified origin. 

We were able to derive the full dependence of $K$ on both $T^A$ and $Q^\alpha,X^i$, by using the standard method 
of working out the reduction of the kinetic terms of the bosonic fields, but only under an a priori strong assumption 
on $X$ and $V$. This assumption consists in some non-trivial properties of the harmonic $1$-forms $u_P$ on $X$ with 
values in $V$, which define the charged matter zero-modes, relative to the harmonic $(1,1)$ forms $\omega_A$ on $X$, 
which define the neutral moduli zero-modes. More precisely, the assumption is that the $(1,1)$ forms 
$c_{PQ} = i\, {\rm tr}(u_P \wedge \bar u_Q)$ are harmonic and can be expanded onto the basis $\omega_A$
with some constant coefficients $c^A_{PQ}$. For models where $X$ and $V$ are such that this is true, $K$ can 
be derived in closed form, with a moduli dependence controlled by the intersection numbers $d_{ABC}$ and a 
matter dependence controlled by the quantities $c^A_{PQ}$, which are constant by assumption. The result that 
we derived precisely matches the general form proposed in \cite{PS} by an $M$-theory argumentation. We however 
believe that its validity is restricted to the situations satisfying the above mentioned assumptions, which we argued to 
be needed also from the $M$-theory viewpoint to be able to safely discard the effect of non-zero modes. 
Unfortunately we have no clear idea on how restrictive the above assumption really is. We however showed 
that compactifications based on orbifolds do automatically satisfy it, as a consequence of the fact that the forms 
$u_P$ and $\omega_A$ are in this case not only harmonic but actually covariantly constant, and explained 
how the known result for $K$ in these models \cite{Korb1,Korb2} emerges from the more general expression
that we derived.

Our main conclusions concerning the possibility of implementing an effective sequestering mechanism based on a 
global symmetry are the following. For simplicity we focused on the reference point corresponding to scalar VEVs 
that are negligible for all the matter fields and sizable only for the moduli fields, where gravitational effects to the 
global symmetry Ward identities trivialize. In the special case of the untwisted sector of singular orbifolds, $d_{abc}$ 
and $c^a_{PQ}$ can be identified with the symmetric invariant symbol and the transposed fundamental representation 
generators of some group $H \subset SU(3)$, and the scalar manifold is a symmetric K\"ahler manifold. 
It then turns out that there exists an exact global symmetry $H \times U(1)^{h^{1,1}-1}$ of $K$ which, if extended 
also to $W$, implies the vanishing of all the contributions to soft terms, with constrained but non-trivial $F^i$ 
although vanishing $F^a$. In the more general case of smooth Calabi-Yau's, on the other hand, $d_{abc}$ 
and $c^a_{PQ}$ have no particular properties, other than being respectively symmetric and Hermitian, and 
the scalar manifold is a generic K\"ahler manifold. 
It then turns out that a similar mechanism can be at work only in the special case where the intersection numbers 
$d_{abc}$ and the matrices $c^a$ are respectively the symmetric invariant and the transposed fundamental 
generators of some group $H$. In such a situation there exists an exact global symmetry 
$H \times U(1)^{h^{1,1}-1}$ of $K$ which, if extended also to $W$, implies the vanishing of all the contributions 
to soft terms, with constrained but non-trivial $F^i$ although vanishing $F^a$.

In summary, it emerges rather clearly that an effective mechanism of sequestering based on a global symmetry 
seems to be naturally possible only whenever the scalar manifold is a very particular space with properties that 
resemble those of symmetric spaces. From an effective theory point of view, the analysis that we have 
done for this presumably larger class of models is then somewhat similar in spirit to the analysis that was 
done in \cite{LPS} for models based on symmetric spaces. More precisely, the authors of \cite{LPS} studied the 
possibility of achieving degenerate boson and fermion masses in some arbitrary sector of the model but at arbitrary 
points by suitably dialing the Goldstino direction, whereas here we studied the possibility of achieving vanishing scalar 
masses in a visible matter sector and at a particular reference point as a robust result of imposing a global symmetry 
on the hidden matter and moduli sector to suitably constrain the Goldstino direction.

\section*{Acknowledgments}

It is a pleasure to thank J.-P.~Derendinger, S.~Ferrara, W.~Lerche, J.~Louis, F.~Paccetti Correia and M.~Schmidt
for useful discussions. This work was supported by the Swiss National Science Foundation.

\appendix

\section{Calabi-Yau manifolds and vector bundles over them}
\setcounter{equation}{0}

In this appendix, we review some notation and results concerning compact Calabi-Yau manifolds $X$ and 
holomorphic vector bundles $V$ over them. We will focus on those results that concern more directly 
$(1,1)$ forms on $X$ and $1$ forms on $X$ with values in $V$, since these are the ingredients that we
need to work out the results we are interested in.

Consider first a compact Calabi-Yau manifold $X$. The tangent and cotangent bundles $TX$ and $T^*X$
have structure group $SU(3)$, since this is the holonomy group characterizing this kind of manifolds. 
We can introduce a basis of $h^{1,1}$ independent harmonic $(1,1)$ forms $\omega_A$ on $X$, which 
provide a basis for the cohomology group $H^{1,1}(X) \simeq H^1(X,T^*X)$:
\be
\{\omega_A\} = \text{basis of $H^{1,1}(X)$} \,.
\ee
We next consider the dual basis of $(2,2)$ harmonic forms $\omega^A$ and the corresponding basis of $4$-cycles 
$\gamma_A$, defined in such a way that
\be
\int_X \!\! \omega_A \wedge \omega^B = \int_{\gamma_A} \!\! \omega^B = \delta_A^B \,.
\ee
We may then define the intersection numbers $d_{ABC}$, which are topological invariants of $X$ counting how many 
times a triplet of $4$ cycles $\gamma^A$, $\gamma^B$ and $\gamma^C$ intersect each other, as
\be
d_{ABC} = \int_X \!\! \omega_A \wedge \omega_B \wedge \omega_C = {\rm intersections} (\gamma_A,\gamma_B, \gamma_C)\,.
\ee
Any harmonic $(1,1)$ form $\sigma$ can be decomposed on the basis $\omega_A$ as 
\be
\sigma = \sigma^A \omega_A \,,
\ee
with real components $\sigma^A$ given by
\be
\sigma^A = \int_X \!\! \omega^A \wedge \sigma \,.
\ee
The Hodge dual $*\sigma$ is a harmonic $(2,2)$ form, and can therefore be decomposed onto the basis of 
$\omega^A$ as
\be
*\sigma = \sigma_A \, \omega^A \,,
\ee
with real components $\sigma_A$ given by
\be
\sigma_A = \int_X \!\! \omega_A \wedge *\sigma \,.
\ee
There always exist at least one harmonic $(1,1)$ form defining the K\"ahler structure:
\be
J = \text{K\"ahler form} \,.
\ee
In fact, it turns out that the volume form $*1$ on $X$ can be expressed as the exterior product 
of three K\"ahler forms $J$:
\be
* 1 = \frac 16 J \wedge J \wedge J \,.
\ee
Integrating this expression over $X$ one deduces that the volume $V$ of $X$ can be
expressed as follows:
\be
V = \frac 16 \int_X \!\! J \wedge J \wedge J \,.
\label{volX}
\ee
As a consequence of the existence and the properties of $J$, the Hodge dual of any harmonic 
$(1,1)$ form $\sigma$ on $X$ can be expressed in the following way in terms of $J$ \cite{Strominger}:
\be
* \sigma =  - J \wedge \sigma + \frac 1{4V} \bigg\{ \int_X \!\! \sigma \wedge J \wedge J \bigg\} \, J \wedge J \,.
\label{relstar}
\ee
In particular, one has:
\be
* J = \frac 12 J \wedge J \,. \label{relstarJ}
\ee
Taking the exterior product of (\ref{relstar}) with any other harmonic $(1,1)$ form $\rho$ and integrating
over $X$, one further deduces that the natural positive-definite scalar product on the space of all the 
harmonic $(1,1)$ forms can be rewritten as:
\be
\int_X \!\! \rho \wedge *\sigma = - \int_X \!\! \rho \wedge \sigma \wedge J 
+ \frac 1{4V} \int_X \!\! \rho \wedge J \wedge J \int_X \!\! \sigma \wedge J \wedge J \,.
\ee
In particular, one finds:
\bea
\a\a \int_X \!\! J \wedge *J = 3 V \,, \label{relstar1} \\
\a\a \int_X \!\! \omega_A \wedge *J = \frac 12 \int_X \!\! \omega_A \wedge J \wedge J \,, \label{relstar2} \\
\a\a \int_X \!\! \omega_A \wedge *\omega_B = - \int_X \!\! \omega_A \wedge \omega_B \wedge J 
+ \frac 1{4V} \int_X \!\! \omega_A \wedge J \wedge J \int_X \!\! \omega_B \wedge J \wedge J \,. \label{relstar3}
\eea
Dividing by $V$ and using the decomposition $J = J^A \omega_A$, which implies that 
$\omega_A = \partial J/ \partial J^A$, these relations can also be rewritten in the following 
more compact form:
\bea
\a\a \frac 1{V} \int_X \!\! J \wedge *J = 3 \,, \label{rel1} \\
\a\a \frac 1{V} \int_X \!\! \omega_A \wedge *J = \frac {\partial}{\partial J^A} \log V \,, \label{rel2} \\
\a\a \frac 1{V} \int_X \!\! \omega_A \wedge *\omega_B = - \frac {\partial^2}{\partial J^A \partial J^B} \log V\,. \label{rel3}
\eea

Consider now a holomorphic vector bundle $V$ over $X$, with structure group $S$. Out of this we can 
define a whole family of vector bundles $V_r$ associated to any representation ${\bf r}$ of $S$, by promoting 
the transition functions of $V$, which are matrices in the fundamental representation of $S$, to the corresponding 
matrices in the representation ${\bf r}$ of $S$. We can then introduce a basis of $n_R$ harmonic $1$-forms $u_P$ 
taking values in the representation ${\bf r}$ of the Lie algebra of $S$, associated to the cohomology group 
$H^1(X,V_{\rm r})$:
\be
\{u_P\} = \text{basis of $H^1(X,V_{\rm r})$} \,. 
\ee
By taking the exterior product of such a $u_P$ with a conjugate $\bar u_Q$ and tracing over the indices of the 
representation ${\bf r}$, one may construct $(1,1)$ forms on the Calabi-Yau manifold $X$, which are however 
generically not harmonic:
\be
c_{PQ} = i\, {\rm tr} \big(u_P \wedge \bar u_Q \big) \,.
\ee
One may then define the following quantities, which are a priori not topological invariants and depend in general 
on the geometry:
\be
c^A_{P Q} = \int_X \!\! \omega^A \wedge c_{P Q} \,,
\ee
In the particular cases where the $(1,1)$ forms $c_{PQ}$ are harmonic, the quantities $c^A_{PQ}$ 
represent their components on the basis defined by the $\omega_A$, and one may then write 
$c_{PQ} = c^A_{PQ} \omega_A$. More in general, one may write a Hodge decomposition 
with exact and coexact terms parametrized by generic $(1,0)$ and $(1,2)$ forms $\alpha_{PQ}$ and $\beta_{PQ}$:
\be
c_{PQ} = c^A_{PQ} \omega_A + \bar \partial \alpha_{PQ} + \bar \partial^\dagger \!\beta_{PQ} \,.
\ee

Notice that by performing general linear transformations one may choose convenient special bases 
$\{\hat \omega_A\}$ and $\{\hat u_P \}$ for harmonic $(1,1)$ forms and Lie-algebra-valued $1$ forms. 
For instance, one may define canonical bases by requiring that the $\hat \omega_A$ and 
$\hat u_P$ should form orthonormal sets with respect to the positive definite scalar products that can be 
defined on them. More precisely, we can impose that 
\bea
\a\a \hat \omega_A :\;\; \frac 1V \int_X \!\! \hat \omega_A \wedge *\hat \omega_B = \delta_{AB} \,, \\
\a\a \hat u_P :\;\; \frac 1V \int_X \!\! \hat c_{PQ} \wedge *J = \delta_{PQ} \,.
\eea
One may moreover orient these bases with respect to the K\"ahler form, in such a way that 
$\hat \omega_0 = J/\sqrt{3}$ and thus $*J = \sqrt{3}\, V \hat \omega^0$. By using eqs.~(\ref{relstar1})-(\ref{relstar3}) 
it then follows that in such a basis the intersection numbers $\hat d_{ABC}$ and the quantities $\hat c^A_{PQ}$ 
have the following special structure:
\bea
\a\a \hat d_{000} = \frac 2{\sqrt{3}} \cdot V \,,\;\; \hat d_{00a} = 0 \cdot V \,,\;\; 
\hat d_{0ab} = - \frac 1{\sqrt{3}} \cdot V \,,\;\; \hat d_{abc} = \text{generic} \cdot V \,, \\
\a\a \hat c^0_{PQ} = \frac 1{\sqrt{3}} \delta_{PQ} \,,\;\; \hat c^a_{PQ} = \text{generic}\,.
\eea

We would like to conclude this appendix by making a few comments concerning the particular case of orbifolds, 
where the Calabi-Yau manifold $X$ degenerates to the projection of a flat torus and the
holomorphic vector bundle $V$ over it is correspondingly constructed as the projection of a trivial bundle. 
In that case, the whole technology simplifies and most of the relations listed above map to simple identities 
in linear algebra. Recall for instance that for any invertible square matrix $M$, the definitions of determinant, 
cofactor and inverse imply that:
\bea
M^{-1}_{ij} \b=\b \frac {\text{cofactor}_{ji} M}{\det M} = \frac {\partial_{M_{ji}} \det M}{\det M} \nn \\[1mm]
\b=\b  \partial_{M_{ji}} \log \det M \,. \label{Minv}
\eea
Moreover, starting from $M_{ik} M^{-1}_{kj} = \delta_{ij}$, taking a derivative and multiplying 
by the inverse, one also deduces that:
\bea
M^{-1}_{ij} M^{-1}_{pq} \b=\b - \partial_{M_{jp}} M^{-1}_{iq} = - \partial_{M_{qi}} M^{-1}_{pj} \nn \\[1mm]
\b=\b - \partial_{M_{jp}} \partial_{M_{qi}} \log \det M \label{MinvMinv} \,.
\eea
Applying these relations to the matrix formed by the components of the metric, one then sees that 
(\ref{Minv}) and (\ref{MinvMinv}) essentially correspond to (\ref{rel2}) and (\ref{rel3}).

\section{Symmetric coset manifolds}
\setcounter{equation}{0}

In this appendix, we summarize some basic facts about the geometry of the symmetric scalar manifolds 
appearing in the low energy effective theories of orbifold compactifications. These have the form 
${\cal M} = {\cal G}/{\cal H}$, where the isometry group ${\cal G}$ is a non-compact Lie group and 
the isotropy group ${\cal H}$ is a maximal compact subgroup of it. Rather than studying separately 
the three kinds of spaces (\ref{M1}), (\ref{M2}) and (\ref{M3}), we shall focus on their basic building block, 
which is the following Grassmannian coset space for $p=1,2,3$ and arbitrary integer $n$, which 
has complex dimension $p(p+n)$:
\be
{\cal M} = \frac {SU(p,p+n)}{U(1) \times SU(p) \times SU(p+n)} \,.
\label{Grass}
\ee

The canonical parametrization of the above space involves a rectangular $p \times (p+n)$ matrix of 
complex coordinates $Z^{iJ}$, with $i=1, \cdots,p$, $s=1,\dots,n$ and $I =i,s$. In this parametrization, 
the full stability group ${\cal H} = U(1) \times SU(p) \times SU(p+n)$ acts linearly on $Z^{iJ}$, 
in the bifundamental representation $({\bf p}, {\bf p+n})_1$. Moreover, at the reference 
point $Z^{iJ} = 0$ these canonical coordinates correspond to normal coordinates, with trivial metric and 
vanishing Christoffel symbols. The K\"ahler potential reads \cite{CV}:
\be
K = - \log \det \big(1 - Z \bar Z \big) \,.
\ee
The parametrization that naturally emerges in the string setting is however a slightly different one. 
It involves a $p \times p$ matrix of moduli coordinates $T^{ij}$ and a $p \times n$ matrix $\Phi^{is}$ 
of matter coordinates. These are related as follows to the $p \times p$ and $p \times n$ sub-blocks 
$Z^{ij}$ and $Z^{is}$ of the above canonical coordinates $Z^{iJ}$:
\be
Z^{ij} = \bigg(\frac {1 - 2\, T}{1 + 2\, T}\bigg)^{\!\!ij} \,,\;\;
Z^{is} = \bigg(\frac {2 \,\Phi}{1 + 2\, T} \bigg)^{\!\!is} \,.
\ee
In this new parametrization, the action of ${\cal H}$ is more complicated. However, the subgroup 
$U(1) \times SU(p)_{\rm diag} \times SU(n) \subset {\cal H}$ still acts linearly on $T^{ij}$, $\Phi^{is}$,
in the adjoint and bifundamental representations $({\bf 1} \oplus {\bf p^2-1},1)_0$ and $({\bf p}, {\bf n})_1$. 
In particular, under the universal subgroup $U(p) \simeq U(1) \times SU(p)_{\rm diag}$ that is independent 
of $n$, $T^{ij}$ and $\Phi^{is}$ transform in the adjoint and the fundamental representations ${\bf n^2}$ 
and ${\bf n}$. 
Moreover, at the reference point $T^{ij} = 1/2\,\delta^{ij}$, $\Phi^{is} = 0$ these new coordinates are only almost
normal coordinates, with trivial metric but some non-vanishing Christoffel symbols. The K\"ahler 
potential becomes, up to a K\"ahler transformation \cite{Korb1}:
\be
K = - \log \det \big(T + \bar T - \Phi \bar \Phi\big) \,.
\ee

The manifold under consideration is not only homogeneous but actually symmetric, since the Lie algebra 
$g$ of ${\cal G}$ is the sum of the Lie algebra $h$ of ${\cal H}$ and a normal component 
$n$ associated to ${\cal G}/{\cal H}$, $g = h \oplus n$, such that $[h,h] \subset h$, $[h,n] \subset n$ and $[n,n] \subset h$. 
This implies that the Riemann curvature tensor is covariantly constant, $\nabla_m R_{i \bar \jmath p \bar q} = 0$.
As a consequence, the metric and the curvature tensors with tangent space indices are both completely fixed in 
terms of group theoretical properties of ${\cal G}$ and ${\cal H}$. To be more precise, let us label the 
generators of $g$ with $T^X$, those of $h$ with $T^x$ and finally those of $n$ with $T^\theta$. The 
metric is then given by the Killing form of $g$ restricted to $n$:
\be
g_{\theta \bar \xi} = - B_{\theta \xi} \,.
\ee
The Riemann tensor is instead fixed by the structure constants ruling the part $[n,n] \subset h$ of the algebra, and reads
\be
R_{\theta \bar \xi \sigma \bar \tau} = f_{\theta \xi}^{\;\;\;\,x} f_{\sigma \tau}^{\;\;\;\,y} B_{xy}  \,.
\ee
Note that although the Killing form $B_{XY}$ on $g$ is indefinite, its restriction $B_{\theta \xi}$ to $h$ is negative 
definite, so that the above metric is positive definite, and its restriction $B_{xy}$ to $n$ is positive definite, 
so that the curvature is negative definite. 

For the manifold at hand, it is a simple exercise to compute the components of the metric and the Riemann 
tensor. To do so, it is convenient to switch to the standard two-index labeling of the generators of unitary groups.
The generators $T^{\Theta\Gamma}$ of $U(p,p+n)$ satisfy  $[T^{\Theta\Gamma},T^{\Sigma \Delta}] = 
\eta^{\Gamma\Sigma} T^{\Theta \Delta} - \eta^{\Theta \Delta} T^{\Gamma\Sigma}$. The generators 
$T^{ij}$ and $T^{IJ}$ of the subgroups $U(p)$ and $U(p+n)$ similarly satisfy  
$[T^{ij},T^{kl}] = \delta^{jk} T^{il} - \delta^{il} T^{jk}$ and  $[T^{IJ},T^{KL}] = -\delta^{JK} T^{IL} + \delta^{IL} T^{JK}$. 
The remaining generators $T^{iJ}$ and $T^{Ij}$ in the coset $U(p,p+n)/(U(p)\times U(p+n))$, which are 
associated to the fields $Z^{iJ}$ and their conjugate $\bar Z^{I \bar \jmath}$, satisfy instead the following 
commutation relations: $[T^{iJ},T^{kL}] = 0$, $[T^{Ij},T^{kL}] = 0$, $[T^{iJ},T^{Kl}] = - \delta^{JK} T^{il} - \delta^{il} T^{JK}$, 
$[T^{Ij},T^{kL}] = \delta^{jk} T^{IL} + \delta^{IL} T^{jk}$. The metric is trivial:
\be
g_{i I \bar \jmath \bar J} = \delta_{ij} \delta_{IJ} \,.
\ee
The Riemann tensor is instead found to be given by the following simple expression,
which can also be verified by a direct computation using canonical coordinates at 
the reference point as in \cite{CV}:
\be
R_{i I \bar \jmath \bar J k K \bar l \bar L} = \delta_{ij} \delta_{kl} \delta_{IL} \delta_{JK} + \delta_{il} \delta_{jk} \delta_{IJ} \delta_{KL} \,.
\ee
Finally, one may split the $p(p+n)$ "complex" coset generators $T^{iJ}$ into moduli generators $T^{im}$ and matter generators 
$T^{i\alpha}$. The metric then splits into
\be
g_{i m \bar \jmath \bar n} = \delta_{ij} \delta_{mn} \,,\;\;
g_{i \alpha \bar \jmath \bar \beta} = \delta_{ij} \delta_{\alpha\beta} \,,\;\;
g_{i m \bar \jmath \bar \beta} = 0 \,,
\ee
and the Riemann tensor decomposes as
\bea
\a\a R_{i m \bar \jmath \bar n k p \bar l \bar q} = \delta_{ij} \delta_{kl} \delta_{mq} \delta_{np} + \delta_{il} \delta_{jk} \delta_{mn} \delta_{pq} \,, \\
\a\a R_{i \alpha \bar \jmath \bar \beta k \gamma \bar l \bar \delta} = \delta_{ij} \delta_{kl} \delta_{\alpha \delta} \delta_{\beta \gamma} 
+ \delta_{il} \delta_{jk} \delta_{\alpha \beta} \delta_{\gamma \delta} \,, \\
\a\a R_{i m \bar \jmath \bar n k \gamma \bar l \bar \delta} = \delta_{il} \delta_{jk} \delta_{mn} \delta_{\gamma \delta} \,.
\eea

At this point, one may apply the above results to the coset spaces (\ref{M1}), (\ref{M2}) and (\ref{M3}) appearing in orbifold
models. The resulting expressions can be rewritten more conveniently by relabeling the generators associated to the moduli 
with a single index. This can be done in parallel for all the three kinds of models by making use of the $3 \times 3$ matrices 
$\lambda^A$ representing $U(1) \times H$ for the relevant subgroup $H \subset SU(3)$. More precisely, $A=0,\cdots,8$ 
for $H = SU(3)$, $a=0,\cdots,3,8$ for $H = SU(2) \times U(1)$ and $a=0,3,8$ for $H = U(1) \times U(1)$. Using the normalization 
condition ${\rm tr}(\lambda^A \lambda^B) = \delta^{AB}$ and the completeness properties applying to each of the three subsets of 
matrices, the metric is found to be
\be
g_{A \bar B} = \delta_{AB} \,,\;\;
g_{i \alpha \bar \jmath \bar \beta} = \delta_{ij} \delta_{\alpha\beta} \,,\;\;
g_{A \bar \jmath} = 0 \,,
\ee
and the Riemann tensor reads
\bea
\a\a R_{A \bar B C \bar D} = {\rm tr} (\lambda^A \lambda^B \lambda^C \lambda^D) 
+ {\rm tr} (\lambda^A \lambda^D \lambda^C \lambda^B)\,, \\[0.5mm]
\a\a R_{i \alpha \bar \jmath \bar \beta k \gamma \bar l \bar \delta} = \lambda^A_{il} \lambda^A_{kj} \delta_{\alpha \delta} \delta_{\beta \gamma}
+ \lambda^A_{ij} \lambda^A_{kl} \delta_{\alpha \beta} \delta_{\gamma \delta} \,, \\
\a\a R_{A \bar B k \gamma \bar l \bar \delta} = (\lambda^B \lambda^A)_{kl} \delta_{\gamma \delta} \,.
\eea

\end{document}